\documentclass[aps,preprint,preprintnumbers,showpacs,amsmath,amssymb,superscriptaddress,footinbib,longbibliography]{revtex4-1}
\usepackage[pdftex, colorlinks, citecolor=blue]{hyperref}   % this is for pdflatex pdf format
\usepackage{graphicx}
\usepackage{dcolumn}
\usepackage{threeparttable}
\usepackage{multirow}
\usepackage{booktabs}
\usepackage{txfonts}
\usepackage{xcolor}
\usepackage{bm}
\usepackage{amssymb}
\usepackage{amsmath}
\usepackage{latexsym}
\usepackage{epsfig}
\usepackage{amsbsy}
\usepackage{array}
\usepackage{tabularx}
\usepackage{esvect} %vector
\usepackage{extarrows}
\usepackage{float}
\usepackage[british]{babel}
\usepackage[section]{placeins}
\usepackage{makecell}

\usepackage[normalem]{ulem}

\begin{document}

% \title{Development of an accurate and efficient moment tensor machine-learning interatomic potential for describing defects in Ni-Al alloys}

\title{Efficient moment tensor machine-learning interatomic potential for accurate description of defects in Ni-Al Alloys}

\author{Jiantao Wang}
\affiliation{%
Shenyang National Laboratory for Materials Science, Institute of Metal Research, Chinese Academy of Sciences, 110016 Shenyang, China
}%
\affiliation{%
School of Materials Science and Engineering, University of Science and Technology of China, 110016 Shenyang, China
}%

\author{Peitao Liu}%
\email{ptliu@imr.ac.cn}
\affiliation{%
Shenyang National Laboratory for Materials Science, Institute of Metal Research, Chinese Academy of Sciences, 110016 Shenyang, China
}%

\author{Heyu Zhu}
\affiliation{%
Shenyang National Laboratory for Materials Science, Institute of Metal Research, Chinese Academy of Sciences, 110016 Shenyang, China
}%

\author{Mingfeng Liu}%
\affiliation{%
Shenyang National Laboratory for Materials Science, Institute of Metal Research, Chinese Academy of Sciences, 110016 Shenyang, China
}%

\author{Hui Ma}%
\affiliation{%
Shenyang National Laboratory for Materials Science, Institute of Metal Research, Chinese Academy of Sciences, 110016 Shenyang, China
}%

\author{Yun Chen}%
\affiliation{%
Shenyang National Laboratory for Materials Science, Institute of Metal Research, Chinese Academy of Sciences, 110016 Shenyang, China
}%

\author{Yan Sun}%
\affiliation{%
Shenyang National Laboratory for Materials Science, Institute of Metal Research, Chinese Academy of Sciences, 110016 Shenyang, China
}%

\author{Xing-Qiu Chen}%
\email{xingqiu.chen@imr.ac.cn}
\affiliation{%
Shenyang National Laboratory for Materials Science, Institute of Metal Research, Chinese Academy of Sciences, 110016 Shenyang, China
}%

\begin{abstract}
 Combining the efficiency of semiempirical potentials with the accuracy of quantum mechanical methods,
machine-learning interatomic potentials (MLIPs) have significantly advanced atomistic modeling in computational materials science and chemistry.
This necessitates the continual development of MLIP models with improved accuracy and efficiency,
which enable long-time scale molecular dynamics simulations to unveil the intricate underlying mechanisms that would otherwise remain elusive.
Among various existing MLIP models,
the moment tensor potential (MTP) model employs a highly descriptive rotationally covariant moment tensor to describe the local atomic environment,
enabling the use of even linear regression for model fitting.
Although the current MTP model has achieved state-of-the-art computational performance for similar accuracy,
there is still room for optimizing the contraction process of moment tensors.
In this paper, we propose an effective genetic algorithm based optimization scheme
that can significantly reduce the number of independent moment tensor components and intermediate tensor components.
This leads to a speedup of nearly one order of magnitude in evaluation and also improved accuracy compared to the traditional MTP model for intricate basis sets.
We have applied our improved MTP model to predicting the energetic and dynamical properties
of various point and planar defects in Ni-Al alloys,
showing overall good performance and in general outperforming the semiempirical potentials.
This paper paves the way for fast and accurate atomistic modeling of complex systems
and provides a useful tool for modeling defects in Ni-Al alloys.
\end{abstract}

\maketitle

\section{Introduction}

Nickel-based high-temperature alloys hold significant importance in engineering applications, particularly in gas turbines and aerospace, owing to their exceptional high-temperature mechanical properties.
The superior mechanical properties originate from the ordered $\gamma^\prime$-Ni$_3$Al phase embedded in the $\gamma$ phase matrix~\cite{gassonSuperalloysFundamentalsApplications2008}. The presence of a lattice misfit leads to the formation of a dislocation network between the $\gamma$ and $\gamma^\prime$ phases, which prevents dislocations from sliding through the $\gamma^\prime$ phase, resulting in increased strength and creep resistance~\cite{raePrimaryCreepSingle2007}.
Moreover, defects such as impurities, vacancies, dislocations, and stacking faults are prevalent in these alloys, significantly influencing their mechanical strength and resistance to creep.
For instance, manipulating complex stacking faults and twinning boundaries has been discovered to effectively enhance the hardening properties of Ni-Al alloys~\cite{zhangHardeningNi3AlComplex2021}.
Furthermore, the aggregation of vacancies in Ni-Al alloys can lead to the formation of vacancy-type dislocation loops, voids, and stacking fault tetrahedra (SFT) under conditions like quenching, irradiation, plastic deformation, and exposure to a hydrogen environment~\cite{kiritaniSimilarityDifferenceFcc2000,kraftmakherEquilibriumVacanciesThermophysical1998,aidhyFormationGrowthStacking2016, chiariFormationTimeDynamics2021}. These defects have a direct impact on the mechanical properties of the alloys. Therefore, accurately predicting the energetic and dynamic properties of these defects in Ni-Al alloys is crucial for optimizing their mechanical performance.

Density functional theory (DFT) has significantly advanced our comprehension of defect
behavior in the Ni-Al alloys~\cite{doi10.1063/1.2051793, Shang_2012, Shang_2012_2, YU20095914,
YU201238, EURICH201587, YANG2020109682, HU2020155799, XIA2022104183, ZHAO2022110990,
zhuComprehensiveInitioStudy2023,tanDislocationCoreStructures2019,nivaccluster1,
xueUnravelingFormationMechanism2021,rubanFirstprinciplesStudyPoint2014,ZHU202354}.
For instance, Siegel~\cite{doi10.1063/1.2051793} reported that alloying elements can reduce the stacking fault energies in Ni.
Shang \emph{et al.}~\cite{Shang_2012} further studied the effects of various alloying elements on stacking fault energies in Ni, demonstrating that most alloying elements decreased the stacking fault energy, with a more pronounced effect observed when the alloying element is further from Ni in the periodic table. Zhu \emph{et al.}~\cite{zhuComprehensiveInitioStudy2023} highlighted the significant strengthening effects of the alloying elements Re, W, and Mo on the Ni and Ni$_3$Al systems, which increased the slip energy barriers of the trailing slip process in the $\gamma$ phase and the leading slip process in the $\gamma^\prime$ phase.
Using a DFT-based flexible boundary condition approach, Tan \emph{et al.}~\cite{tanDislocationCoreStructures2019} investigated the core structure of $\frac{1}{2}[110]$ screw dislocation in the $\gamma$ phase and $[110]$ screw super dislocation in the $\gamma^\prime$ phase, determining the split distances of planar faults in the dislocation core structure, in good agreement with experimental observations.
Zhao \emph{et al.}~\cite{nivaccluster1} investigated the stability of vacancy clusters in Ni using \emph{ab initio} molecular dynamics (AIMD), revealing that the SFT was the most thermodynamically stable configuration for large vacancy clusters in Ni.

Despite the extensive DFT investigations on the Ni-Al alloys, accurately modeling defects remains a challenging task. On the one hand,
a large simulation cell is often required to eliminate the artificial interactions between the defect and its periodic replicas, thereby substantially increasing the computational expenses. This necessity becomes especially pronounced for intricate defects like vacancy clusters, dislocations, and stacking faults.
On the other hand,
capturing the dynamical properties of defects
necessitates a long timescale that is usually beyond the reach of AIMD simulations.
To overcome the limitations of DFT, several semiempirical interatomic potentials have been developed for the Ni-Al binary system.
Among these potentials, the most successful ones for Ni are the
embedded atom method (EAM) potential developed by Voter \emph{et al.} in 1986~\cite{voterAccurateInteratomicPotentials1986},
Mishin \emph{et al.} in 2004~\cite{mishinAtomisticModelingPhases2004} and 2009~\cite{purjapunDevelopmentInteratomicPotential2009},
and the EAM potential optimized for NiAl and Ni$_3$Al systems developed by Wang \emph{et al.}~\cite{wangEmbeddedatomPotentialNiAl2018}. Besides, Kumar \emph{et al.}~\cite{Kumar_2015} developed a third-generation charge-optimized many-body potential, which was featured by its accurate description of planar defects within the $\gamma$ and $\gamma^\prime$ phase as well as their interfaces with Al$_2$O$_3$.
Notably, Avik \emph{et al.}~\cite{mahataModifiedEmbeddedatomMethod2022}
recently developed a modified EAM potential for the Ni-Al system
that worked in a wide temperature range and can accurately reproduce experimental melting points across all compositions.
Although these physics-based semiempirical potentials were capable of reproducing the target properties used in the fitting such as experimental lattice parameters, elastic constants, cohesive energy, and vacancy formation energy, they often suffered from limited accuracy and failed to capture the properties of complex defects over a large phase space, because of their simple and inflexible analytical forms.

Machine-learning interatomic potentials (MLIPs) have emerged as an elegant solution to bridge the expensive DFT calculations and less accurate semiempirical potentials~\cite{NNP1,artrithHighdimensionalNeuralnetworkPotentials2011, doi:10.1021/acs.chemrev.0c00868,hanDeepPotentialGeneral2018, wangDeePMDkitDeepLearning2018,
Marques_npj2019, PhysRevLett.124.086102,GAP1,GAPdoctorthisis,vaspmlotf,jinnouchiPhaseTransitionsHybrid2019,
10.1063/1674-0068/cjcp2108145,PhysRevB.105.L060102,
thompsonSpectralNeighborAnalysis2015,mtp1,mtpal1,
drautzAtomicClusterExpansion2019,lysogorskiyPerformantImplementationAtomic2021a, dussonAtomicClusterExpansion2022,
10.1063/5.0166858,10.1063/5.0106617,
doi:10.1021/acs.chemrev.1c00022,doi:10.1021/acs.chemrev.0c00868,doi:10.1021/acs.chemrev.0c01111,
doi:10.1021/acs.chemrev.1c00021, doi:10.1021/acs.chemrev.0c01303,
Yanxon_2021,Sauceda2022,Batzner2022,Allegro_2023,XiangHongjun_2024,Sunjian_2024,XIE2024,MACE2024,DPA2_2024}.
These MLIPs utilize machine learning techniques to fit the high-fidelity quantum mechanical data,
thereby maintaining the high efficiency of semiempirical potentials while preserving quantum mechanical accuracy.
In recent years, the MLIPs have been successfully applied to metals and alloys~\cite{PhysRevB.90.104108,PhysRevMaterials.2.013808,Maresca2018,Kostiuchenko2019,Vandermause2020,
PhysRevB.100.144105,PhysRevMaterials.4.093802,Nikoulis_2021,PhysRevMaterials.5.053804,PhysRevB.103.094112,PhysRevMaterials.7.045802,
Hart2021,PITIKE2023154183,PhysRevMaterials.7.043601,PhysRevB.108.054312,Wen_2022,PhysRevMaterials.7.103602,Liu2024,Owen2024,Cheng2024}.
With variations in the construction of the training set, structural descriptors, and regression techniques,
several MLIP models have been proposed.
Behler and Parrinello~\cite{NNP1} pioneered the locality assumption and proposed a high-dimensional neural network potential (HDNNP) model.
As of now, the HDNNP model has advanced to the fourth generation, incorporating long-range Coulomb interactions, charge transfer, and magnetic moments~\cite{doi:10.1021/acs.chemrev.0c00868,Ko2021}.
The deep potential model~\cite{hanDeepPotentialGeneral2018, wangDeePMDkitDeepLearning2018} also employed a deep neural network but using an embedding network of atomic coordinates as local configuration descriptors,
and has achieved great success in various systems~\cite{Wen_2022,doi:10.1126/science.ado1744,Niu2020,Yang2023,Zhu2024}.
Another popular and successful MLIP model is the kernel-based Gaussian approximation potential (GAP) proposed by Bart\'{o}k~\emph{et al.}~\cite{GAP1,GAPdoctorthisis},
which employed the descriptor of the smooth overlap of atomic positions (SOAP) and used Gaussian process regression~\cite{GAP1,GAPdoctorthisis}, demonstrating promising results in many applications~\cite{doi:10.1021/acs.chemrev.1c00022}.
Jinnouchi \emph{et al.}~\cite{vaspmlotf,jinnouchiPhaseTransitionsHybrid2019} further developed the GAP model
by incorporating an on-the-fly active learning procedure within AIMD simulations using the Bayesian force errors, effectively addressing the laborious  process of conventional training set construction.
Despite the popularity of the GAP potential, the kernel-based models suffer efficiency issues, especially as the training set grows large.
Instead of employing complex neural networks or kernel functions, linearly parameterized models
such as the spectral neighbor analysis potential (SNAP)~\cite{thompsonSpectralNeighborAnalysis2015}, moment tensor potential (MTP)~\cite{mtp1, mtpal1},
and atomic cluster expansion (ACE)~\cite{drautzAtomicClusterExpansion2019, dussonAtomicClusterExpansion2022, lysogorskiyPerformantImplementationAtomic2021a}
have been developed owing to their highly descriptive nature. Zuo \emph{et al.}~\cite{zuoPerformanceCostAssessment2020} systematically assessed the accuracy and efficiency of the HDNNP, GAP, MTP, SNAP, and quadratic SNAP (qSNAP)~\cite{woodExtendingAccuracySNAP2018} models on various lattice types,
demonstrating that among these tested models, the MTP potential exhibits the highest efficiency for the same level of accuracy.

While the MTP model has reached state-of-the-art levels in efficiency and accuracy, there remains room for improvement.
In this paper, we advanced the MTP model by developing an elegant optimization scheme for the contraction of moment tensors.
Thanks to the utilization of lower-rank moment tensors and more efficient contraction rules,
the number of independent moment tensor components and intermediate tensor components in our improved MTP model was significantly reduced.
This resulted in a nearly tenfold increase in speed and enhanced accuracy as compared to the traditional MTP model for the basis sets with high level of complexity. As a benchmark application of our improved MTP model, we have systematically assessed its performance in describing the energetic and dynamical properties of various point and planar defects in Ni-Al alloys, an established challenging task for MLIPs. Our results showcased consistently good performance of our developed MTP model, in general outperforming the semiempirical EAM potentials.
This was manifested by accurately predicting various properties such as the lattice constants, elastic properties, energy-volume curves, and phonon dispersions as well as the melting points of bulk Ni and Ni$_3$Al.
Furthermore, our MTP model accurately predicted the formation and binding energies of vacancy clusters, antisite defects, stacking faults, and the diffusion behavior of vacancy clusters.
This study paves the way for fast and accurate atomistic modeling of complex systems and lays the foundation for further improving the current MTP model for Ni-Al alloys by expanding the phase space of the training set.

The paper is organized as follows. In Sec.~\ref{sec:method} we will
detail the methodology of our improved MTP model.
Particular emphasis is devoted to the
proposed optimization scheme for contracting moment tensors and basis set selection.
Technical details and computational setups will be provided in Sec.~\ref{sec:details}.
The extensive results will be presented and discussed in
Sec.~\ref{sec:results}, followed by the conclusion in Sec.~\ref{sec:conlcusions}.

\section{Method development}\label{sec:method}

\subsection{Moment tensor potential}\label{sec:mtmmodel}

We start from the introduction of the moment tensor potential (MTP) proposed by Shapeev~\cite{mtp1}. The key quantity of the MTP is the rotationally covariant moment tensor descriptor~\cite{mtp1}
\begin{equation}\label{eq:Moment_tensor}
	M_{\mu, \nu}\left(\mathfrak{n}_{i}\right)=\sum_{j} f_{\mu,\nu}\left(\left|\mathbf{r}_{i j}\right|, z_{i}, z_{j}\right) f_c(\left| \mathbf{r}_{i j} \right|) \underbrace{\mathbf{r}_{i j} \otimes \ldots \otimes \mathbf{r}_{i j}}_{\nu \text { times }}~,
\end{equation}
which describes the local atomic environment $\mathfrak{n}_{i}$ of atom $i$. The moment tensor $M_{\mu,\nu}$ is a tensor of rank $\nu$. $f_{\mu, \nu}$ denotes radial basis functions (e.g., Chebyshev polynomials), $z_i$ denotes the species index of atom $i$, and $f_c$ represents the cutoff function that approaches zero smoothly at a specific cutoff radius $r_c$. $\mathbf{r}_{ij}$ represents the distance vector from atom $i$ to atom $j$, and the symbol $\otimes$ denotes the Kronecker product.

It is worth noting that the rank of the moment tensor corresponds to the rank of the momenta of the weighted atom mass density~\cite{mtp1}. Physically, one can think of the moment tensor descriptors as analogues to mass distribution properties. For instance, the rank-0 moment tensor descriptor simply gives the total mass and the rank-1 moment tensor descriptor describes the center of mass. The rank-2 moment tensor descriptor represents the second moments of inertia,
while among higher-order moment tensors, the rank-3 tensor can be specifically interpreted as a statistical measure of skewness, which quantifies the asymmetry of the mass distribution.
% while for higher orders of moment tensors, like rank-3 moment tensor, can be interpreted as the statistical measure of skewness, which quantifies the asymmetry of the mass distribution.

By appropriately contracting a set of $k$ moment tensors ${ M_{\mu_i,\nu_i} }$, one is able to obtain a scalar-valued $k$-body function~\cite{mtp1}. Such contraction can be represented by a symmetric index matrix $\alpha$ of size $k\times k$ with diagonal terms being zero. The $i$th row of the matrix $\alpha$ corresponds to the moment tensor $M_{\mu_i,\nu_i}$, and the sum of the elements in the $i$th row of the matrix $\alpha$ is equal to $\nu_i$, (i.e., $\nu_i=\sum_j \alpha_{ij}$). The off-diagonal element $\alpha_{ij}$ describes how many dimensions are contracted between the moment tensors $M_{\mu_i,\nu_i}$ and $M_{\mu_j,\nu_j}$. Using this contraction representation, the scalar-valued function can be described as
$B_\alpha={\prod\limits_{i}}^\alpha M_{\mu_i,\nu_i}$~\cite{mtp1}.
To better illustrate the contraction, here we present an example. Suppose that we have three moment tensors $M_{1, 3}$, $M_{2, 2}$, and $M_{3, 1}$ [i.e., $\mu=(1, 2, 3)^T$ and $\nu=(3, 2, 1)^T$]  and a contraction rule
$
	\alpha =
	\left(
	\begin{array}{c c c}
		0 & 2 & 1 \\
		2 & 0 & 0 \\
		1 & 0 & 0
	\end{array}
	\right)
$, then the final contracted scalar function $B_\alpha$ can be represented in Einstein summation rule as $B_\alpha =(M_{1, 3})_{ijk} (M_{2, 2})_{ij} (M_{3, 1})_{k}$.

The scalar basis functions $\{B_\alpha\}$ are invariant under translation, rotation, and reflection. It has been shown by Shapeev~\cite{mtp1} that all these scalar functions form a complete basis set.
Therefore, the local atomic energy $U_i$ of atom $i$ can be written as linear combinations of these scalar basis functions
\begin{equation}
U_i = \sum_\theta \xi_\theta B_\theta,
\end{equation}
where $\xi_\theta$ are linear coefficients. The summation over all atoms then yields the potential energy of the system $U = \sum_i U_i$.

\subsection{Optimization of moment tensors contraction}\label{sec:contraction_opt}

\begin{figure}[H]
	\begin{center}
		\includegraphics[width=0.8\textwidth]{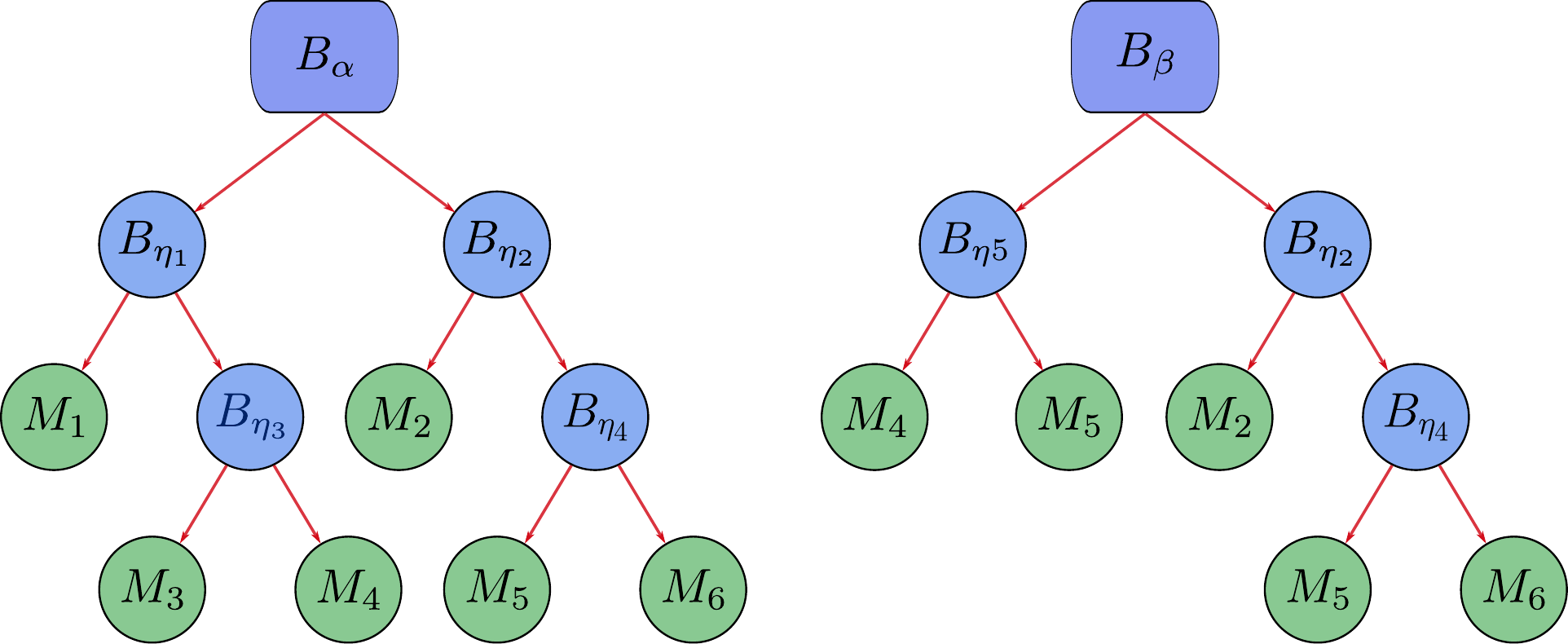}
	\end{center}
	\caption{Illustration of the tree representation of scalar decomposition. $B_\alpha$ and $B_\beta$ are two different scalar functions, $B_{\eta_i}$ are indeterminate tensors, and $M_i$ are moment tensors. Note the common indeterminate tensors $B_{\eta_2}$ and $B_{\eta_4}$ between the two different scalar functions.
	}
	\label{fig:decomp}
\end{figure}

In practical implementation, directly computing the contractions is computationally demanding, since the entries to be summed grow exponentially with the count of contracted indices (namely, $3^m$ when contracting $m$ indices between two moment tensors).
Thanks to the symmetric nature of the moment tensors, this computational cost can be reduced to $(m+2)(m+1)/2$ when incorporating symmetry of the moment tensors. To reduce the cost further,
Shapeev~\cite{mtp1} found an ingenious scheme. In this scheme, as illustrated in Fig.~\ref{fig:decomp}, a contraction process can be represented as a tree. Starting from a scalar function that is defined by contractions of moment tensors, a tensor is decomposed into contraction of two intermediate tensors iteratively until moment tensors are reached. This process actually defines the order of contractions and represents the intermediate results. It can be seen that decomposition for different scalar functions may contain the same intermediate tensors. Therefore, if all these common intermediate tensors are computed once and used for all, the entire tensor contraction cost can be further reduced.

To illustrate this procedure in detail, the indices of a tensor $B_\alpha$ that is decomposed into a tensor $B_\gamma$ of rank $k_\gamma$ and a tensor $B_\eta$ of rank $k_\eta$ can be expressed as
\begin{align}
	& \alpha = \left( \begin{array}{c c}
		\gamma^\prime & \phi \\
		\phi^T & \eta^\prime
	\end{array} \right), \\
	& \gamma_{i,i} = \gamma^\prime_{i,i} + \sum_j \phi_{i, j}, \quad \eta_{i,i} =  \eta^\prime_{i,i} + \sum_j \phi^T_{i, j}, \\
	& \gamma_{i,j}=\gamma^\prime_{i,j}, \quad \eta_{i,j}=\eta^\prime_{i,j}, \quad  (i\neq j),
\end{align}
where $\gamma^\prime$ and $\eta^\prime$ are the diagonal blocks of the index matrix $\alpha$ with sizes of $k_\gamma \times k_\gamma$ and $k_\eta \times k_\eta$, respectively; $\gamma$ and $\eta$ are index matrices for tensors $B_\gamma$ and $B_\eta$, respectively; and $\phi$ is the off-diagonal block of $\alpha$ with size $k_\gamma \times k_\eta$.
The nonzero diagonal terms $\alpha_{ii}$ of nonscalar index $\alpha$ denote that the $\alpha_{ii}$ indices of moment tensor $M_i$ are not involved in contraction, whereas the off-diagonal terms $\alpha_{ij}$ represent contracting $\alpha_{ij}$ indices between moment tensors $M_i$ and $M_j$. The nonzero elements $\phi_{i, j}$ of matrix $\phi$ represent contracting $\phi_{i,j}$ indices between the indices of $B_\gamma$ inherited from moment tensor $M_i$ and indices of $B_\eta$ inherited from moment tensor $M_j$.
The indices $\alpha$, $\gamma$, $\eta$, and $\phi$ define a contraction rule to obtain $B_\alpha$ from $B_\gamma$ and $B_\eta$.

Based on the above analysis, the computation routine for the moment tensors contraction in practice can then be established. First, the components $\{m_{i,j}\}$ of moment tensor $M_i$ can be calculated directly from atomic coordinates. The components of intermediate tensor $B_{\eta_3}$ in Fig.~\ref{fig:decomp}, for example, can be calculated by $b_{\eta_3,i} = \sum_k m_{3,k} m_{4,k}$ where $k$ loops over contracted indices, and so do other intermediate tensors. Taking $B_i$ as root node and $B_l$ and $B_m$ as leaf nodes for instance, the components of $B_i$ can be expressed as $b_{i, j} = \sum_k b_{l, k} b_{m, k}$. By assigning each nonequivalent tensor component with a unique ID, each intermediate tensor component and scalar tensor component can be evaluated through a collection of tuples $\{(n,d_i,k_i,l_i)\}$
\begin{align}
\label{eq:evalbs}
b_n = \sum_i d_i \cdot b_{k_i} \cdot b_{l_i},
\end{align}
where $d_i$ is the degeneracy owing to tensor symmetry, $n$, $k_i$, and $l_i$ are unique IDs of tensor components. We refer to these tuples as \textit{times rules}. Once these rules have been built, one can employ them at any time without rebuilding them. Additionally, the count of nonequivalent moment tensor elements ($n_m$), the count of nonequivalent intermediate tensor elements ($n_b$), and the count of rule applications ($n_t$) can all be determined. From Eq.~\eqref{eq:evalbs}, it is evident that the computational cost for evaluating all tensor elements, or equivalently, performing all tensor contractions, is proportional to $n_t$.

It needs to be noted that for a specific scalar function $B_\alpha$, the decomposition routine is not unique.
For instance, in the case presented in Sec.~\ref{sec:mtmmodel}, the scalar function $B_\alpha$ can be obtained by three different contraction processes: $(M_{1,3}:M_{2,2})\cdot M_{3,1}$, $(M_{1,3}\cdot M_{3,1}):M_{2,2}$ and $(M_{2,2}\otimes M_{3,1})_{ijk} (M_{1,3})_{ijk}$.
In fact, for a collection of scalar functions $\{ B_\alpha \}$, different choices of decomposition routines for each scalar function can lead to varied numbers of $n_b$ and $n_t$. This, in turn, significantly influences the computational cost of the model training and applications. Because of the large number of possible decomposition routines, it is practically impossible to loop over all cases.
For instance, in the case of the moment tensors with level 28 as defined by Shapeev~\cite{mtpal2},
there are more than $10^{2195}$ possible decomposition routines,
which are derived by multiplying the number of unique decompositions for each scalar function in the basis set defined by the level 28.
Shapeev~\cite{mtp1} simply pointed out that minimizing the sum of dimensions for each decomposition and maximizing tensor reuse can serve as a guideline for the selection of decomposition routines.
Although this scheme may result in a reasonable solution,
it could introduce ambiguity, e.g., when multiple decomposition routines have the same sum of dimensions or intermediate tensors, and thus might not consistently ensure the optimal solution.

\begin{figure}[H]
	\begin{center}
		\includegraphics[width=0.8\textwidth]{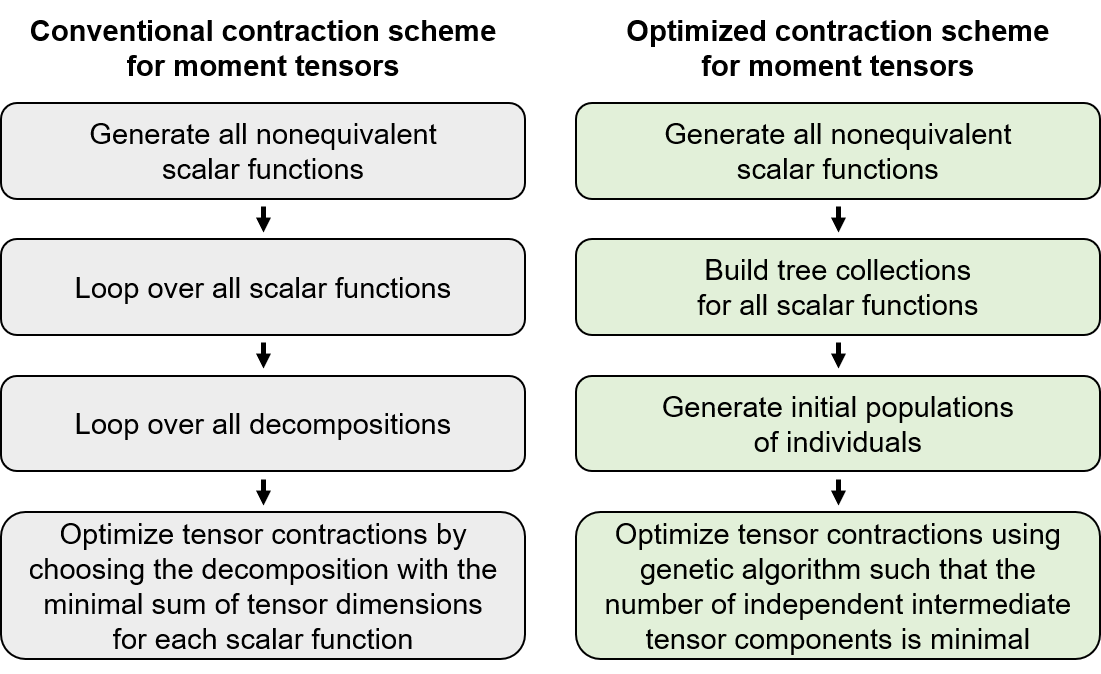}
	\end{center}
	\caption{Schematic workflows illustrating the differences between the conventional (left) and the optimized (right) contraction schemes for moment tensors.
	}
	\label{fig:flow}
\end{figure}

To tackle this issue, we propose a straightforward approach to obtain the optimal decomposition routine for each scalar function using genetic algorithms (GA). Specifically, we construct all possible decomposition trees for each scalar function. These trees constitute a ``tree collection" and are denoted as $\{ T_i^\alpha \}$ for a scalar function $B_\alpha$.
For each scalar function $B_\alpha$, we assign a ``gene" $k$, which picks the $k$-th tree out of $\{ T_i^\alpha \}$. For a collection of scalar functions, a sequence $ \{ k_1, k_2, \cdots, k_n \} $ forms an individual, which represents a decomposition routine. The objective of optimizing the computation routine amounts to minimizing the number of nonequivalent tensor components and the count of times rules. To quantify this, we define a cost function $F(R)$
\begin{equation}
	F(R) = m + 3 t,
\end{equation}
where $R$ represents a specific routine, $m$ is the number of nonequivalent tensor components, and $t$ is the count of times rules. Adding more weight (i.e.,``3") to the count of times rules is because of the fact that this part is
computationally more costly for a relatively large number of basis sets. With the definition of $F(R)$,
the fitness function of each individual can then be defined as
\begin{equation}
	\label{equ:fitness}
	f(R) = 1 - \frac{1}{1+e^{-k\big(F(R) - F(R)_{\rm min}\big) + c}}.
\end{equation}
Here, $F(R)_{\rm min}$ represents the current minimum of $F(R)$ and it will be changed dynamically during the optimization process. The parameters $k$ and $c$ are positive constants.
This functional form is reminiscent of the Fermi-Dirac distribution, where the parameter $k$ regulates the steepness of the decline in fitness values,
while the parameter $c$ shifts the curve along the $F(R)$ axis.
By tuning these parameters, we can effectively exclude solutions that are far from optimal, while preserving those that are closer to the optimal solution within the population.
This approach helps prevent the population from being dominated by identical solutions.
The detailed workflow can be described as follows:
\begin{itemize}
	\item [(1)] Generate all nonequivalent scalar functions according to given conditions (i.e., maximum "level", order of many-body interactions, maximum coordination power, etc.) and then sort these scalar functions by the count of moment tensors involved in contraction.
	
	\item [(2)] Build tree collections for all scalar functions.
	
	\item [(3)] Generate initial populations of individuals. In this step, a score is assigned to each tree in the collection based on specific criteria. For example, a tree may gain ten points if it has the minimum sum of dimensions among all possible decompositions, with one point for each reuse of intermediate tensor and an additional point if a tensor can be decomposed into two identical tensors. Those trees with the highest scores are selected as the initial guesses for the decomposition routine.
	
	\item [(4)] Generate child populations via mutations and crossover operations. The mutation operator randomly selects a scalar function and replaces its current tree with another one from the collection, and the cross-over operator substitutes a subset of an individual's trees with those from another individual.
	
	\item [(5)] Calculate the fitness function of each individual and decide which individuals will survive according to their fitness.
	
	\item [(6)] Repeat steps (4) and (5) until either the designated maximum number of iterations has been attained or the convergence criteria have been reached.	
\end{itemize}
To illustrate the differences between the conventional contraction scheme (i.e., Shapeev's scheme~\cite{mtp1}) and our optimized contraction scheme for moment tensors,
we have created a flowchart, as depicted in Fig.~\ref{fig:flow}.

It should be mentioned that during the search process, we enforce that each tensor is assigned to one decomposition routine. This means that the genes representing the decomposition trees are not independent from each other. In other words, choosing a specific tree for a previously processed scalar function may render the tree structure invalid for a later processed scalar function if they assign the same tensor with nonequivalent decompositions. This will introduce conflicts during the optimization process. To tackle this problem, when generating times rules for the later processed tree, the branches forked from previously processed tensors are chopped so that they can be consistent with the previously processed tensors without changing the genes.
After the optimization using the genetic algorithms, a greedy algorithm search is then carried out to further optimize the computation routine. In this step, all trees are looped from low rank to higher rank and the decomposition routine with the lowest loss function is eventually selected.
We note that while the genetic and greedy algorithms were also utilized in the seminal study of Shapeev~\cite{mtp1},
the primary emphasis of that study lies in the basis set sparsification, which fundamentally differs from those of our paper.
In our paper, these algorithms are applied to optimize the tensor contraction orders within a predefined basis set,
aiming to minimize the number of independent intermediate tensor components and, consequently, the overall computational cost.

\subsection{Basis set selection}

After the rules of moment tensors contraction have been established, a subsequent question is which basis functions should be selected.
To this end, Shapeev \emph{et al.}~\cite{mtpal2} introduced a concept of ``level'' for each moment tensor $M_{\mu,\nu}$
by $\text{lev}M_{\mu,\nu} = 4\mu + \nu + 2$. For a scalar basis $B_\alpha$ obtained from a set of moment tensors $\{M_{\mu_1,\nu_1}, \cdots, M_{\mu_n,\nu_n}\}$,
its level is defined as $\text{lev}B_\alpha = \sum_i^n \text{lev}M_{\mu_i,\nu_i}$~\cite{mtpal2}.
A predefined level threshold $d$ is then used to filter the basis functions satisfying $\text{lev}B_\alpha \leq d$~\cite{novikovMLIPPackageMoment2021}.
Here, we redefine the level of moment tensor as $\text{lev}M_{\mu, \nu} = 2\mu + \nu + 1$.
We made such a choice based on the consideration that the computational complexity of the MTP model primarily arises from two sources: (i) the calculation of the moment tensor components, and (ii) the computation of all tensor components through pregenerated contraction rules.
For a moment tensor of order $\nu$, it has $(\nu+1)(\nu+2)/2$ unique tensor components.
To compute these moment tensor components, one must loop over all $n_i$ neighboring atoms of a central atom $i$.
As $n_i$ scales cubically with the cutoff radius $r_c$, the computational cost of this process thus exhibits a quadratic scaling with the moment tensor order $\nu$ and a cubic scaling with $r_c$.
This part of the computation cannot be optimized and its cost may become dominant as $\nu$ increases. Therefore, we limit the upper bound of $\nu$ to a smaller value ($\nu$$\leq$3) to reduce the number of moment tensor components. Additionally, we retain the complexity of the basis functions by decreasing the penalty factor of $\mu$ from 4 to 2. Our tests showed that this scheme can greatly enhance the computational speed without compromising accuracy. To achieve higher-order polynomial terms, one can introduce scalar basis functions generated by the contraction of a larger number of moment tensors (representing higher-order many-body interactions).

\section{Computational details}\label{sec:details}

\subsection{Training set generation}\label{sec:train}
The training set generation consisted of two processes.
The first process employed the on-the-fly active learning procedure based on the kernel-based Bayesian regression during AIMD simulations as implemented in
the Vienna \emph{Ab initio} Simulation Package (VASP)~\cite{jinnouchiPhaseTransitionsHybrid2019,vaspmlotf}.
The cutoff radius for the descriptors and the width of the Gaussian functions used for broadening the atomic
distributions of the descriptors were set to 5 \AA and 0.5 \AA, respectively.
The number of radial basis functions was set to 10 and 8 for the two-body and three-body descriptors, respectively.

It is recognized that the kernel-based methods can encounter performance issues when the training set grows large,
resulting in a large number of basis functions~\cite{zuoPerformanceCostAssessment2020,liuCombiningMachineLearning2023}.
To address this issue, a more efficient linearly parameterized MTP potential~\cite{mtp1} was fitted on the training set generated in the first process.
When building the MTP basis functions, the maximum number of moment tensors involved in contraction was set to four,
which resulted in up to five-body interactions considered. The maximum level of scalar functions was set to 26 and eight radial basis functions were used.

In the second process, multiple active learning cycles were performed based on Shapeev's generalized D-optimality criterion~\cite{mtpal1} to further sample the phase space and improve the transferability of the model. These cycles began by generating initial configurations that span a broad range of diversity, including various defects.
Subsequently, we carried out 200 ps MD simulations and identified configurations with an extrapolation grade exceeding five.
The DFT energies, forces, and stress tensors of these selected configurations were then computed and  incorporated into the training set at the end of each active learning cycle, followed by a refitting of the MTP model. This process was repeated until no configurations reached the given threshold of extrapolation grade.

\subsection{Basis set selection details}

Following the scheme we proposed in Sec.~\ref{sec:contraction_opt}, here we aim to find the optimal basis functions that yield both accurate and efficient MLIPs by examining five basis sets.
Among them, three basis sets exhibit levels of 18, 24, and 28, as defined in the scheme of Shapeev~\cite{novikovMLIPPackageMoment2021}.
The remaining two basis sets are generated using our proposed scheme, which are referred to as the 2653 and 4613 basis sets. The 2653 basis set considered up to five-body interactions and moment tensors of rank 4, leading to 2653 linear parameters, whereas the 4613 basis set accounted for up to seven-body interactions and moment tensors of rank 6, leading to 4613 linear parameters.
For each basis set, 100,000 GA search iterations were performed. For all the basis sets, the parameter $c$ in Eq.~\eqref{equ:fitness} was set to be 0.2 and the parameter $k$ in Eq.~\eqref{equ:fitness} was set to $5 \times 10^{-2}$ for the level-18 basis set, $5\times 10^{-3}$ for the level-24 and 2653 basis sets, and $5 \times 10^{-4}$ for the level-28 basis set, respectively. The population size was set to 40. At each iteration, 12 individuals were selected by roulette wheel selection to generate new individuals.

\subsection{Additional computational settings}

All first-principles calculations and on-the-fly training set selection were performed using VASP~\cite{vasp1,vasp2}. The Perdew-Burke-Ernzerhof (PBE) exchange-correlation functional~\cite{ggapbe} within the projected augmented wave method~\cite{paw1,pawvasp} was employed. A plane-wave cutoff of 400 eV
and a Monkhorst-Pack $k$-point grid with a spacing of $\sim$0.2 \AA$^{-1}$ were used, which ensures the convergence of total energy within 1 meV/atom. The convergence criteria for the electronic self-consistent calculation and structure optimization were set to $10^{-6}$ eV and 0.01 eV/\AA, respectively.

The phonon dispersions were calculated by the frozen phonon method using the Phonopy code~\cite{togoFirstprinciplesPhononCalculations2023,togoImplementationStrategiesPhonopy2023}. For these calculations, a $3\times3\times3$ supercell and a finite displacement of 0.01~\AA were employed.

Molecular dynamics simulations were conducted using the LAMMPS code~\cite{thompsonLAMMPSFlexibleSimulation2022} in the NPT ensemble at ambient pressure. The Langevin thermostat~\cite{allenThermostatMolecularDynamics2007} and Parrinello-Rahman method~\cite{parrinelloCrystalStructurePair1980,parrinelloPolymorphicTransitionsSingle1981} were used to control the temperature and pressure of the system. The time step was set to 2 fs.
The on-the-fly AIMD samplings were performed by heating the structures from 700 to 1600 K at ambient pressure in an isothermal-isobaric ensemble using a Langevin thermostat~\cite{allenThermostatMolecularDynamics2007}
combined with the Parrinello-Rahman method~\cite{parrinelloCrystalStructurePair1980,parrinelloPolymorphicTransitionsSingle1981}.

\section{Results and discussion}\label{sec:results}

\subsection{Training and validation sets}

\begin{figure}[H]
	\begin{center}
		\includegraphics[width=0.9\textwidth]{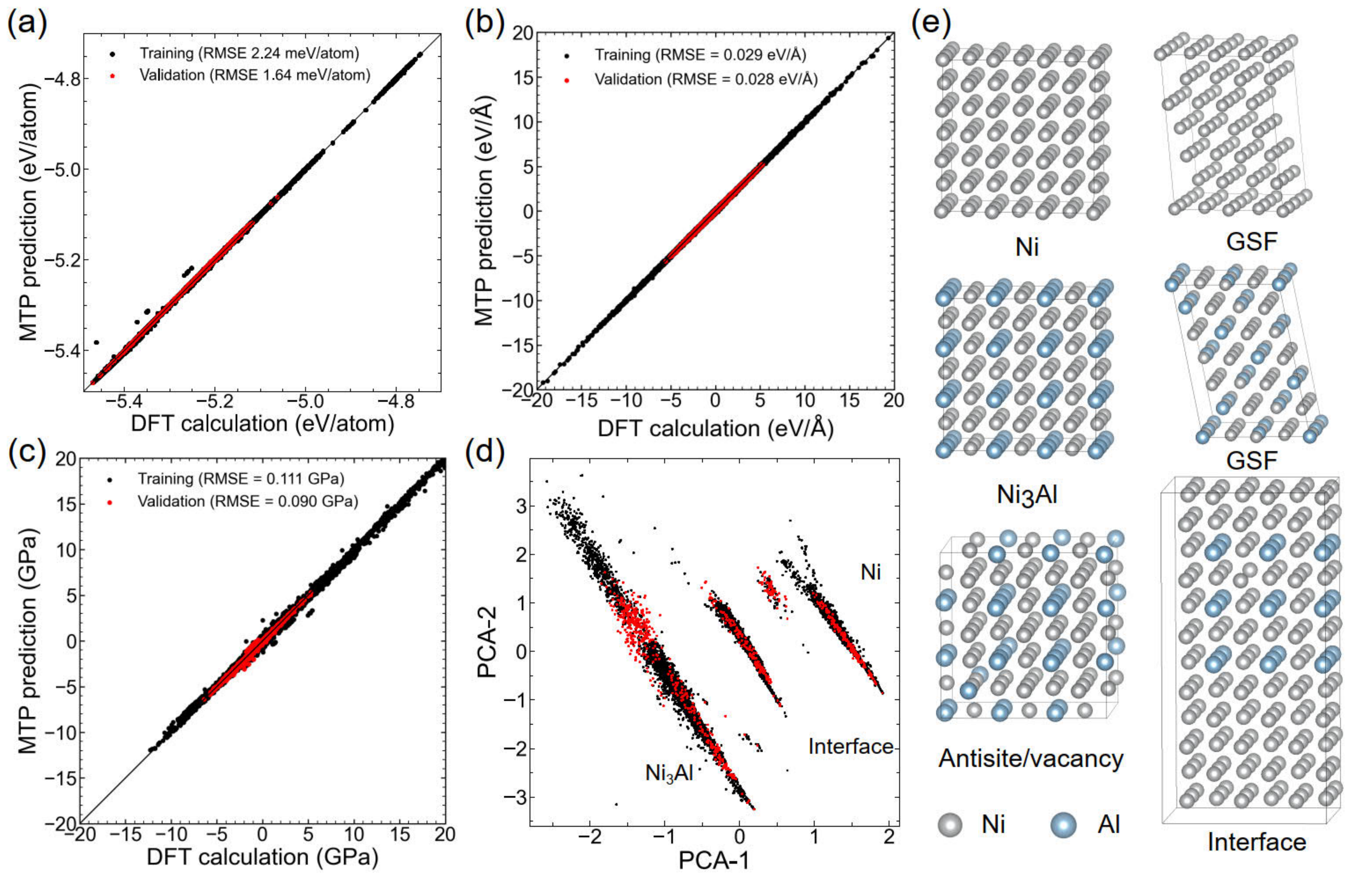}
	\end{center}
	\caption{
		(a)-(c) Comparison of DFT and MTP predicted energies, forces and stresses, respectively.
		(d), PCA analysis of training and validation datasets.
		Black and red points represent training and validation data, respectively.
		(e), Representative structures in the training set.
	}
	\label{fig:asap}
\end{figure}

A dataset consisting of 6092 configurations was obtained in the first process of the sampling procedure described in Sec.~\ref{sec:train}, and
2358 new configurations were collected in the second process, leading to the expansion of the training set to 8450 configurations, on which the final MTP potential was trained.
The dataset consists of fcc Ni, L1$_2$-Ni$_3$Al, Ni/Ni$_3$Al interface, GSF of Ni and Ni$_3$Al, as well as their combinations with vacancies, vacancy clusters, and antisite defects [see Fig.~\ref{fig:asap}(e)]. To assess the accuracy of our developed MLIP model,
a validation set consisting of 815 configurations was constructed using the same structure prototypes as those in the training set.

The training and validation datasets were analyzed by principal component analysis (PCA) using the smooth overlap of atomic positions as local structure descriptors~\cite{chengMappingMaterialsMolecules2020}. The PCA analysis, as shown in Fig.~\ref{fig:asap}(d), demonstrated that the validation set effectively covers the training set. Three distinct clusters were identified,
which correspond to Ni-based, Ni$_3$Al-based, and interface configurations. The Ni$_3$Al-based configurations spread in a larger area, which is caused by the diffusion of vacancies and vacancy clusters leading to antisite defects and defect complexes.

\subsection{Efficiency and accuracy assessments of MLIPs}

Table~\ref{tab:mtm} presents the assessments of different basis sets on the accuracy and efficiency of the resulting MLIPs. One can observe that the optimization process
does not affect the accuracy, but it reduces the number of intermediate tensors, the number of nonequivalent tensor components, as well as the count of times rules for all the basis sets considered.
This reduction becomes more pronounced for more complex basis sets.
For example, for the simple basis set such as level-18, the reduction percentages are 8.0\% for the number of intermediate tensors, 19.2\% for the number of nonequivalent tensor components, and 19.0\% for the count of times rules.
By contrast, for more complex basis sets like level-28, which have more intricate decomposition routines, the optimization process exhibits more significant reductions in the numbers of intermediate tensors (18.2\%), nonequivalent tensor components (41.3\%), and times rules (43.9\%).

\begin{table}[H]
	\begin{center}
		\caption{Assessments of different basis sets on the accuracy and efficiency of MLIPs.
  The columns with ``\#Tensors", ``\#Components", and ``\#Times rules" represent the number of intermediate tensors, the number of nonequivalent tensor components, and the count of times rules, respectively.
  The columns with ``Energies", ``Forces", and ``Stresses" indicate the training (upper row) and validation (lower row) RMSEs of energies (in meV/atom), forces (in meV/\AA), and stress tensors (in kbar), respectively.
  MD times were measured on a $8\times8\times8$ supercell of bulk Ni$_3$Al
  (2048 atoms) with 100,000 time steps using 96 CPU cores [Intel (R)Xeon(R) Platinum 9242 CPU @ 2.30 GHz]  and averaged for five independent runs.
  The column with ``Efficiency" denotes the speedup of the MLIPs with optimized basis sets (OPT) as compared to those with non-optimized (INI) basis sets.
  Note that for these tests all the MLIPs were fitted on the same reduced training dataset (4800 structures) using a reduced cutoff radius of 5~\AA.
  }
		\label{tab:mtm}
		\begin{tabular}{c c | c c c c c c c c }
  \hline
			Basis set & {} & \#Tensors & \#Components & \#Times rules & Energies & Forces & Stresses & MD time (s) & Efficiency  \\
			\hline
			\multirow{4}{2.0em}{18}
			   & \multirow{2}{*}{INI}  & \multirow{2}{*}{225}   & \multirow{2}{*}{736}   & \multirow{2}{*}{2799}   & 3.64 & 0.045 & 2.22 & \multirow{2}{*}{151}   & \multirow{2}{*}{1.00}  \\
		    {} & {}                    & {}                     & {}                     & {}                      & 2.40 & 0.042 & 1.52 & {}                     & {}    \\
			{} & \multirow{2}{*}{OPT}  & \multirow{2}{*}{205}   & \multirow{2}{*}{595}   & \multirow{2}{*}{2254}   & 3.64 & 0.045 & 2.22 & \multirow{2}{*}{145}   & \multirow{2}{*}{1.04}  \\
			{} & {}                    & {}                     & {}                     & {}                      & 2.40 & 0.042 & 1.52 & {}                     & {}    \\
			\hline
			\multirow{4}{2.0em}{24}
			   & \multirow{2}{*}{INI}  & \multirow{2}{*}{1222}  & \multirow{2}{*}{4991}  & \multirow{2}{*}{38970}  & 3.00 & 0.035 & 1.52 & \multirow{2}{*}{917}   & \multirow{2}{*}{1.00}  \\
     		{} & {}                    & {}                     & {}                     & {}                      & 1.98 & 0.033 & 1.11 & {}                     & {}    \\
			{} & \multirow{2}{*}{OPT}  & \multirow{2}{*}{1038}  & \multirow{2}{*}{3403}  & \multirow{2}{*}{25732}  & 3.00 & 0.035 & 1.52 & \multirow{2}{*}{711}   & \multirow{2}{*}{1.29}  \\
			{} & {}                    & {}                     & {}                     & {}                      & 1.98 & 0.033 & 1.11 & {}                     & {}    \\
			\hline
			\multirow{4}{2.0em}{28}
			   & \multirow{2}{*}{INI}  & \multirow{2}{*}{3501}  & \multirow{2}{*}{16504} & \multirow{2}{*}{186495} & 2.43 & 0.030 & 1.24 & \multirow{2}{*}{4418}  & \multirow{2}{*}{1.00}  \\
			{} & {}                    & {}                     & {}                     & {}                      & 1.53 & 0.028 & 0.97 & {}                     & {}    \\
			{} & \multirow{2}{*}{OPT}  & \multirow{2}{*}{2863}  & \multirow{2}{*}{9683}  & \multirow{2}{*}{104664} & 2.43 & 0.030 & 1.24 & \multirow{2}{*}{2580}  & \multirow{2}{*}{1.71}  \\
			{} & {}                    & {}                     & {}                     & {}                      & 1.53 & 0.028 & 0.97 & {}                     & {}    \\
			\hline
			\multirow{4}{2.0em}{2653}
			   & \multirow{2}{*}{INI}  & \multirow{2}{*}{2969}  & \multirow{2}{*}{4764}  & \multirow{2}{*}{15157}  & 1.89 & 0.029 & 1.20 & \multirow{2}{*}{274}   & \multirow{2}{*}{1.00}  \\
			{} & {}                    & {}                     & {}                     & {}                      & 1.38 & 0.028 & 0.92 & {}                     & {}    \\
			{} & \multirow{2}{*}{OPT}  & \multirow{2}{*}{2854}  & \multirow{2}{*}{3822}  & \multirow{2}{*}{11581}  & 1.89 & 0.029 & 1.20 & \multirow{2}{*}{241}   & \multirow{2}{*}{1.14}  \\
			{} & {}                    & {}                     & {}                     & {}                      & 1.38 & 0.028 & 0.92 & {}                     & {}    \\
			\hline
			\multirow{4}{2.0em}{4613}
			   & \multirow{2}{*}{INI}  & \multirow{2}{*}{5264}  & \multirow{2}{*}{7676}  & \multirow{2}{*}{18634}  & 1.62 & 0.026 & 1.02 & \multirow{2}{*}{474}   & \multirow{2}{*}{1.00}  \\
			{} & {}                    & {}                     & {}                     & {}                      & 1.11 & 0.026 & 0.88 & {}                     & {}    \\
			{} & \multirow{2}{*}{OPT}  & \multirow{2}{*}{4964}  & \multirow{2}{*}{6140}  & \multirow{2}{*}{13322}  & 1.62 & 0.026 & 1.02 & \multirow{2}{*}{403}   & \multirow{2}{*}{1.17}  \\
			{} & {}                    & {}                     & {}                     & {}                      & 1.11 & 0.026 & 0.88 & {}                     & {}    \\
			\hline
		\end{tabular}
	\end{center}
\end{table}

\begin{table}[H]
	\begin{center}
		\caption{The training costs of the non-optimized (INI) and optimized (OPT) MLIPs for different basis sets. The training costs were evaluated using 96 CPU cores. The training costs for linear fitting were averaged over five independent runs, while the training costs for each L-BFGS iteration were averaged over 1000 iterations. All the settings are the same as in Table~\ref{tab:mtm}.}
		\label{tab:train}
		\begin{tabular}{c c | c c c c}
			\hline
			\makecell{Basis set} & {} & \makecell{L-BFGS\\ (second)} & \makecell{L-BFGS\\ efficiency} & \makecell{Linear fitting\\ (second)} & \makecell{Linear fitting\\ efficiency} \\
			\hline
			\multirow{2}{2.0em}{18} & INI & 6.52 & \multirow{2}{*}{1.07} & 18 & \multirow{2}{*}{1.28} \\
			                        & OPT & 6.07 & {}                    & 13 & {} \\
			\hline
			\multirow{2}{2.0em}{24} & INI & 36.61& \multirow{2}{*}{1.20} &100 & \multirow{2}{*}{1.08} \\
									& OPT & 29.13& {}                    & 92 & {} \\
			\hline
			\multirow{2}{2.0em}{28} & INI &134.54& \multirow{2}{*}{1.38} &615 & \multirow{2}{*}{1.09} \\
			                        & OPT &83.98 & {}                    &559 & {} \\
			\hline
			\multirow{2}{2.0em}{2653}& INI & 12.74 & \multirow{2}{*}{1.15} & 599 & \multirow{2}{*}{1.02} \\
			                        & OPT  & 10.82 & {}                    & 589 & {} \\
			\hline
			\multirow{2}{2.0em}{4613}& INI & 21.49  & \multirow{2}{*}{1.16} & {1697} & \multirow{2}{*}{1.00} \\
			                        & OPT  & 18.01  & {}                    & {1694} & {} \\
			\hline
		\end{tabular}
	\end{center}
\end{table}

As for the efficiency,
although the optimization process reduces
both the number of tensor components and times rules by 19\% for the level-18 basis set,
it only slightly decreases the MD time cost (3\%).
This is because in this case the computation cost is primarily driven by calculating moment tensor components rather than contractions.
However, when it comes to the more complex basis sets such as level-28, the optimized MLIP exhibits a notable 71\% acceleration. This is because of the fact that for complex bases containing higher order many-body interactions and more intricate contraction operations, loops over times rules and computing derivatives associated with tensor components begin to dominate the computing cost.
As a result, in this case the optimization process of contractions can significantly enhance the overall computational efficiency.

Next, we turn to evaluating the performances of the 2653 and 4613 basis sets that were generated using our proposed scheme. As compared to the level-28 basis set (with 2445 scalar functions), the 2653 basis set contains a similar number of scalar functions and fitting coefficients.
However, the trained MLIP using the 2653 basis set is nearly 11 times faster than using the level-28 basis set. Furthermore, the former exhibits a higher accuracy (see Table~\ref{tab:mtm}).
The enhanced efficiency of the 2653 basis set can be attributed to the utilization of lower-rank moment tensors and more efficient contraction rules.
This results in a notable decrease in the number of independent moment tensor components (scaling quadratically with the tensor rank) and intermediate tensor components.
It is worth mentioning that theoretically, the level-28 basis set is expected to achieve superior accuracy owing to its higher polynomial order (22 compared to 12 for the 2653 basis set excluding the radial function component). However, the comparatively reduced accuracy of the level-28 basis set, as evidenced in our findings, may be attributed to the heightened complexity of the basis set, which complicates the optimization process.

Table~\ref{tab:train} summarizes the training costs of both non-optimized and optimized MLIPs across different basis sets.
Overall, the optimized MLIPs require lower training costs compared to their non-optimized counterparts.
The efficiency improvements observed during L-BFGS iterations generally align with those in MD simulations, except for the level-28 basis set.
Specifically, for the level-28 basis set, the training cost is reduced by 38\%, which is lower than the 71\% reduction observed in MD simulations.
For linear fitting, the reduction in training cost due to optimization is negligible for all basis sets except level-18.
This is primarily because the computational cost is dominated by solving the linear system rather than evaluating MLIPs.
However, for the level-18 basis set, where the size of the linear problem is relatively small, the efficiency gains from optimizing the decomposition routine are thus more pronounced.

While the 4613 basis set offers the highest accuracy, in this study, the 2653 basis set is ultimately chosen for its notable efficiency and relatively high accuracy.
With the 2653 basis set, the MLIP was refitted
on the full training dataset (8450 structures) and a larger cutoff of 5.4~\AA.
To achieve greater accuracy of the MLIP, we implemented a two-step final fitting procedure.
The first step was to obtain suitable initial linear coefficients. This was achieved by the following processes.
A minimal basis set was initially extracted from the full basis set, utilizing the same radial basis functions.
This minimal basis set was utilized to fit the training set.
Subsequently, the initial linear coefficients
were obtained by conducting a linear optimization
on the full basis set, inheriting the radial basis functions of the potential.
The weights of energies, forces, and stress tensors
used in fitting were set to 1, 0.01, and 0.005, respectively.
The second step involved optimizing the fitting parameters.
This was executed using the L-BFGS algorithm with 5000 iterations, followed by a least-squares linear fit to further refine the linear parameters.

The accuracy of the ultimately fitted MLIP was assessed against the DFT calculations. Figure~\ref{fig:asap} illustrates the comparison between DFT and MTP-predicted energies, forces, and stresses on both the training and validation sets, demonstrating outstanding agreement. The root mean square errors (RMSEs) of the energies, forces, and stresses for the training set are 2.24 meV/atom, 0.029 eV/\AA$^2$, and 0.11 GPa, whereas for the validation set the RMSEs are 1.64 meV/atom, 0.028 eV/\AA$^2$, and 0.09 GPa, respectively. To conclude this section, we have developed an efficient and accurate MTP model for the Ni-Al systems.

\subsection{Assessment of MLIP for predicting fundamental physical properties}

We further assessed the performance of our developed MLIP on various physical properties of Ni and Ni$_3$Al systems. For comparison purposes, we also employed the most widely used EAM-type semiempirical interatomic potential for the Ni-Al alloy developed by Mishin~\cite{purjapunDevelopmentInteratomicPotential2009}.
We first computed the lattice parameters, elastic constants, bulk modules, shear moduli and Yong's moduli
as well as Poisson's ratio of both bulk Ni and Ni$_3$Al. The results are compared to the predictions from DFT and EAM methods as well as the experimental data, as summarized in Table~\ref{tab:tconstants}.
One can see that the MTP-predicted lattice parameters
accurately reproduce the DFT results.
The underestimation in the lattice parameters compared to experimental data can be attributed to two factors: (i) The experimental data were obtained at room temperature, and (ii) more importantly, the current MLIP does not incorporate magnetic effects.
Furthermore, the MTP reproduces well the elastic properties compared to the DFT results---both are in good agreement with the experimential data.
Regarding the elastic properties, the EAMs also demonstrate good performance. This outcome is not so surprising as the employed semiempirical EAM potential was derived through fitting specifically to the elastic properties~\cite{purjapunDevelopmentInteratomicPotential2009}.

\begin{table}[H]
	\begin{center}
		\caption{Lattice constant $a$ (in \AA), elastic constants ($C_{11}$, $C_{12}$, and $C_{44}$, in GPa), elastic moduli (bulk modulus $B$, shear modulus $G$, and Young's modulus $E$, in GPa), and Poisson's ratio $v$
  of bulk Ni and Ni$_3$Al predicted by DFT, MTP, and EAM methods. The available experimental (EXP) values are given for comparison.}
		\label{tab:tconstants}
        \begin{threeparttable}
		\begin{tabular}{c c | c c c c c c c c}
  \hline
			{} & {} & $a$ & $C_{11}$ & $C_{12}$ & $C_{44}$ & $B$ & $G$ & $E$ & $v$ \\
			\hline
			\multirow{4}{2.5em}{\rm Ni}
			   & DFT & 3.511 & 257.53 & 171.26 & 117.70 & 200.02 & 78.73 & 208.79 & 0.326 \\
			{} & MTP & 3.508 & 238.71 & 176.04 & 113.09 & 196.93 & 67.86 & 182.61 & 0.345 \\
			{} & EAM & 3.520 & 241.34 & 150.85 & 127.34 & 181.01 & 84.15 & 218.57 & 0.299 \\
			{} & EXP & 3.524~\cite{arblasterSelectedValuesCrystallographic2018} & 249$\pm$4$^a$    & 155$\pm$7$^a$    & 114$\pm$12$^a$    & 190$\pm$13$^a$ & 78$\pm$5$^a$ & 197$\pm$15$^a$ & 0.296$\pm$ 0.029$^a$ \\		

   \hline
			\multirow{5}{2.5em}{Ni$_3$Al}
			   & DFT & 3.562 & 237.56 & 156.40 & 125.80 & 182.86 & 80.27 & 210.08 & 0.309 \\
			{} & MTP & 3.563 & 236.50 & 155.81 & 117.98 & 182.71 & 76.80 & 202.08 & 0.316 \\
			{} & EAM & 3.533 & 237.32 & 166.38 & 130.16 & 190.03 & 77.61 & 204.94 & 0.320 \\
			\multirow{2}{2.5em}{} & \multirow{2}{2.5em}{EXP} &
				3.567~\cite{latexpt1}  & 230~\cite{elasticexp1} & 150~\cite{elasticexp1} & 131~\cite{elasticexp1} & {} & {} & {} & {} \\
			{} & {} &	{} &  221~\cite{koiwaElasticConstantsIntermetallic1994} & 146~\cite{koiwaElasticConstantsIntermetallic1994} & 124~\cite{koiwaElasticConstantsIntermetallic1994} & 171~\cite{koiwaElasticConstantsIntermetallic1994} & 77.8~\cite{prikhodkoTemperatureCompositionDependence1999} & 203.1~\cite{prikhodkoTemperatureCompositionDependence1999} & 0.398~\cite{koiwaElasticConstantsIntermetallic1994}  \\
			\hline
		\end{tabular}
        \begin{tablenotes}
        \item[a] Reference ~\cite{ledbetterElasticPropertiesMetals1973}
        \end{tablenotes}
        \end{threeparttable}
	\end{center}
\end{table}

\begin{figure}[H]
	\begin{center}
		\includegraphics[width=0.9\textwidth]{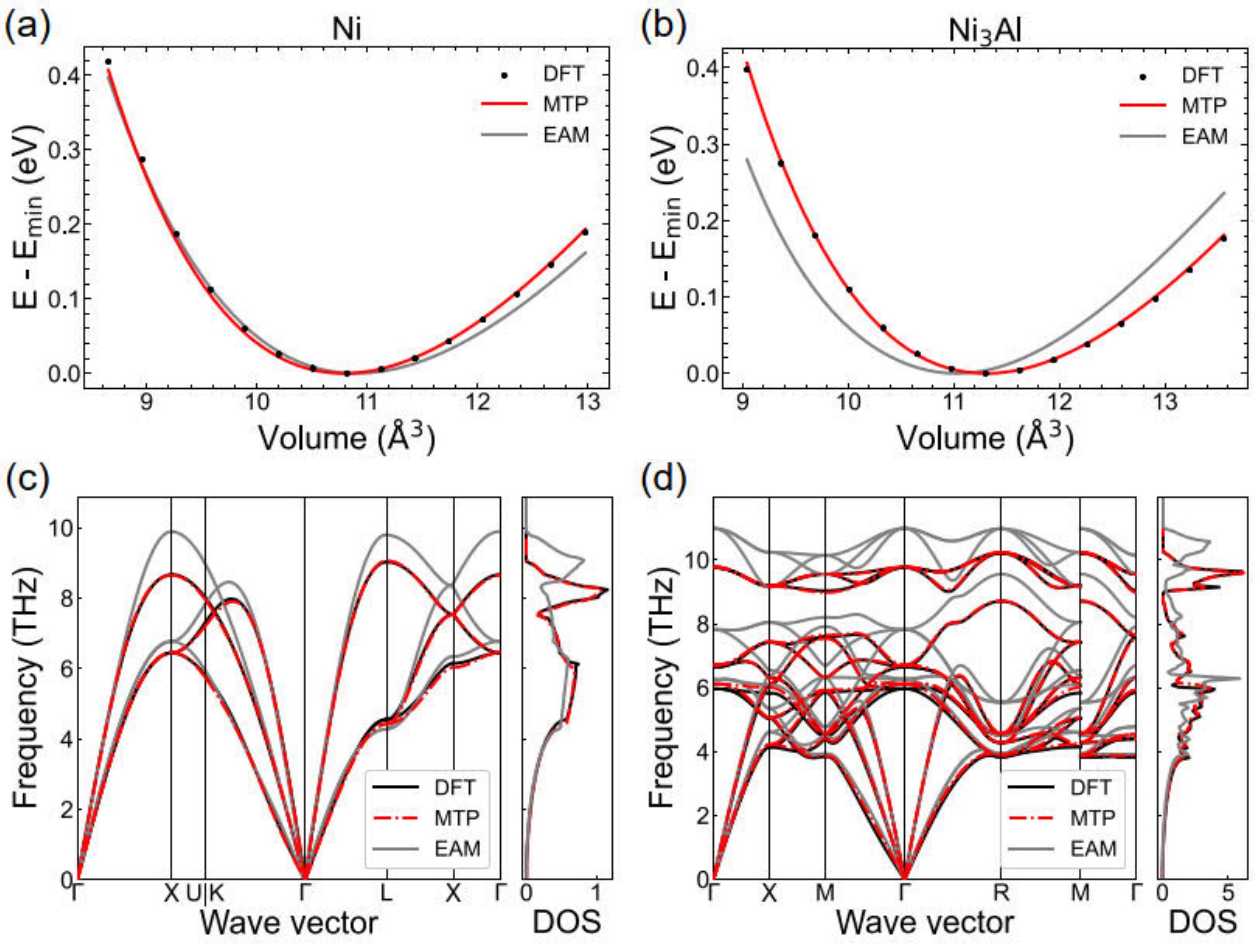}
	\end{center}
	\caption{(a)(b), The energy-volume curves of bulk Ni and Ni$_3$Al predicted by DFT, MTP, and EAM methods.
 (c)(d) The phonon dispersion relationships and phonon density of states of bulk Ni and Ni$_3$Al predicted by DFT, MTP, and EAM methods.
	}
	\label{fig:bulk}
\end{figure}

Figure~\ref{fig:bulk} displays the predicted energy-volume curves, phonon dispersion relations, and phonon density of states
of bulk Ni and Ni$_3$Al by DFT, MTP, and EAM methods.
In comparison to the EAM potential, the developed MTP potential accurately reproduces the DFT-calculated energy-volume curves across all the volume ranges under consideration.
By contrast, the EAM potential diverges from the DFT outcomes as the system moves away from the equilibrium volume.
Regarding the phonon properties, our developed MTP potential demonstrates exceptional alignment with the reference DFT results, once more surpassing the EAM potential.

\subsection{Description of the energetics of point defects by MLIP}

Since a crucial goal of our developed MLIP is to provide an accurate description of point defects, including vacancies and vacancy clusters, within the Ni-Al system,
this section extensively evaluated the MLIP on various types of point defects, including monovacancies, divacancies, trivacancies, vacancy clusters, and antisite defects.

The vacancy formation energy for monovacancy and vacancy clusters in bulk Ni is defined as
\begin{equation}
	E_f(n V_{\rm Ni}) = E(n V_{\rm Ni}) - (N-n) E_{\rm pristine}/N,
\end{equation}
where $E(n V_{\rm Ni})$ indicates the total energy of the defective system containing $n$ vacancies $V_{\rm Ni}$, $E_{\rm pristine}$ stands for the total energy of the pristine bulk Ni system with $N$ atoms. In bulk Ni$_3$Al and Ni-Ni$_3$Al interface systems, the
vacancy formation energy for monovacancy and vacancy clusters
is defined as~\cite{zhangFirstprinciplesStudyVacancy2009}
\begin{equation}
	E_f({n_{\rm Ni}V_{\rm Ni},n_{\rm Al}V_{\rm Al}}) = E_{defect} - E_{prefect} + n_{\rm Ni} E_{ref}^{\rm Ni} + n_{\rm Al} E_{ref}^{\rm Al} + n_{\rm Ni} \mu_{\rm Ni} + n_{\rm Al} \mu_{\rm Al},
\end{equation}
where $E_{defect}$ represents the total energy of the defective system with $n_{\rm Ni}$ Ni vacancies and $n_{\rm Al}$ Al vacancies, $E_{perfect}$ denotes the total energy of the defect-free system, while $E_{ref}^{\rm Ni}$ and $E_{ref}^{\rm Al}$ are the reference energies of bulk Ni and bulk Al in a face-centered cubic cell, respectively,
and $\mu_{\rm Ni}$ and $\mu_{\rm Al}$ stand for the chemical potentials of the Ni and Al atoms in bulk Ni$_3$Al, respectively.
Here, a Ni-rich reservoir was considered, meaning that $\mu_{\rm Ni}$=0 eV and $\mu_{\rm Al}$ is equal to the chemical potential of bulk Ni$_3$Al.

The binding energy of vacancy clusters is calculated as
\begin{equation}
	E_b({n_{\rm Ni}V_{\rm Ni},n_{\rm Al}V_{\rm Al}}) = n_{\rm Ni} E_{V_{\rm Ni}} + n_{\rm Al} E_{V_{\rm Al}} - E_{defect} - (n_{\rm Ni} + n_{\rm Al} - 1) E_{prefect},
\end{equation}	
where $E_{defect}$ represents the total energy of the defective system with $n_{\rm Ni}$ Ni vacancies and $n_{\rm Al}$ Al vacancies, $E_{perfect}$ denotes the total energy of the defect-free system, $E_{V_{\rm Ni}}$ and $E_{V_{\rm Al}}$ denote the total energies of the systems with a Ni vacancy and an Al vacancy, respectively.
Following this convention, positive and negative binding energies denote the attractive and repulsive interactions of the vacancy clusters, respectively.

The antisite defect Ni$_{\rm Al}$ denotes a defective system in which a Ni atom occupies an Al site, while the antisite defect Al$_{\rm Ni}$ signifies a defective system where an Al atom occupies a Ni site.
An antisite pair (Al$_{\rm Ni}$,Ni$_{\rm Al}$) is formed by exchanging the positions of a Ni and an Al atom.
The formation energy of an antisite defect or an antisite pair is simply calculated as $E_f({\rm antisite})= E_{defect} - E_{prefect}$,
where $E_{defect}$ and $E_{prefect}$ are the total energies of the systems with and without antisite defects, respectively.

\begin{table}[H]
	\begin{center}
		\caption{Formation and binding energies of point defects and clusters in bulk Ni and Ni$_3$Al predicted by MTP, DFT, and EAM methods. The available experimental results are given for comparison. $1NN$, $2NN$, and $3NN$ represent the first-, second-, and third-nearest neighbors.
All energies are given in eV.
  }
		\label{tab:pointdefects}
		\begin{tabular}{c c | c c c | c}
  \hline
			{} & {} & MTP & DFT & EAM &  Experiments  \\
			\hline
			\multirow{5}{2.5em}{\rm Ni}
		       & $E_f$($V_{\rm Ni}$)         & 1.425   & 1.425    & 1.572 &  1.4$\pm$0.2~\cite{glazkovFormationPointDefects1987}, 1.74$\pm$0.06~\cite{nanaoStudiesDefectsThermal1977}, 1.73$\pm$0.07~\cite{wolffVacancyFormationNickel1997}  \\
			{} & $E_m$($V_{\rm Ni}$)        & 1.000   & 1.000    & 1.195 &  1.24~\cite{iskakovDeterminationVacancyMigration2015}, 1.38~\cite{girifalcoVacanciesInterstitialsMetals1971}     \\
			{} & $E_f^{di-1NN}$ & 2.835   & 2.799    & 2.890  & 2.92$\sim$3.10~\cite{nanaoStudiesDefectsThermal1977}, 2.42~\cite{seegerAnalysisNonlinearArrhenius1967}, 2.46~\cite{schulePropertiesVacanciesDivacancies1979}   \\
			{} & $E_f^{di-2NN}$($V_{\rm Ni}$,$V_{\rm Ni}$) & 2.924   & 2.912    & 3.129  & {}   \\
			{} & $E_f^{di-3NN}$($V_{\rm Ni}$,$V_{\rm Ni}$) & 2.853   & 2.865    & 3.141  & {}   \\
			{} & $E_b^{di-1NN}$($V_{\rm Ni}$,$V_{\rm Ni}$) & 0.015   & 0.051    & 0.164  & 0.4$\pm$0.2~\cite{nanaoStudiesDefectsThermal1977}, 0.28~\cite{seegerAnalysisNonlinearArrhenius1967}, 0.44~\cite{schulePropertiesVacanciesDivacancies1979}   \\
			{} & $E_b^{di-2NN}$($V_{\rm Ni}$,$V_{\rm Ni}$) &$-$0.074 &$-$0.063  & 0.016  & {}   \\
			{} & $E_b^{di-3NN}$($V_{\rm Ni}$,$V_{\rm Ni}$) &$-$0.003 &$-$0.016  & 0.004  & {}   \\
			\hline
			\multirow{5}{2.5em}{Ni$_3$Al}
		       & $E_f$($V_{\rm Ni}$)              & 1.588 & 1.643  & 1.773 & 1.6$\pm$0.2~\cite{wangStudyVacanciesIntermetallic1984a}   \\
			{} & $E_f$($V_{\rm Al}$)             & 1.710 & 1.787  & 1.544 & 1.6$\pm$0.2~\cite{wangStudyVacanciesIntermetallic1984a}    \\
			{} & $E_b^{di-1NN}$($V_{\rm Ni}$,$V_{\rm Ni}$)   &$-$0.121 &$-$0.154 &$-$0.121 & {}   \\
			{} & $E_b^{di-2NN}$($V_{\rm Ni}$,$V_{\rm Ni}$)   & 0.068 & 0.107 &$-$0.009 & {}   \\
			{} & $E_b^{di-3NN}$($V_{\rm Ni}$,$V_{\rm Ni}$)   & 0.115 & 0.125 &$-$0.005 & {}   \\
			{} & $E_b^{di-1NN}$($V_{\rm Ni}$,$V_{\rm Al}$)   &$-$0.048 &$-$0.076 &$-$0.179 & {}   \\
			{} & $E_b^{di-2NN}$($V_{\rm Ni}$,$V_{\rm Al}$)   &$-$0.044 &$-$0.043 &$-$0.008 & {}   \\
			{} & $E_b^{di-1NN}$($V_{\rm Al}$,$V_{\rm Al}$)   & 0.031 & 0.053 & 0.000 & {}   \\
			{} & $E_b^{di-2NN}$($V_{\rm Al}$,$V_{\rm Al}$)   & 0.004 & 0.002 & 0.019 & {}   \\
			{} & $E_f$(Al$_{\rm Ni}$) & 0.957 & 1.081 & 1.436 & {}    \\ % Al sub Ni site
			{} & $E_f$(Ni$_{\rm Al}$) & 0.159 & 0.033 &$-$0.168  & {}   \\ % Ni sub Al site
			{} & $E_f^{1NN}$(Al$_{\rm Ni}$,Ni$_{\rm Al}$) & 0.842 & 0.839 & 0.901 & {}   \\
			{} & $E_f^{2NN}$(Al$_{\rm Ni}$,Ni$_{\rm Al}$) & 1.098 & 1.080 & 1.237 & {}   \\
			{} & $E_f^{3NN}$(Al$_{\rm Ni}$,Ni$_{\rm Al}$) & 1.120 & 1.095 & 1.256 & {}   \\
			\hline
		\end{tabular}
	\end{center}
\end{table}

All the formation and binding energies in the present study were calculated using a $3\times3\times3$ supercell of 108 atoms. The minimal energy path (MEP) and migration energy barrier $E_m$ of monovacancy were calculated using the climbing image nudged elastic band (CI-NEB) method~\cite{henkelmanClimbingImageNudged2000}.

Table~\ref{tab:pointdefects} summarizes the calculated
formation and binding energies of point defects and clusters in bulk Ni and Ni$_3$Al predicted by MTP, DFT, and EAM methods.
One can observe that our developed MTP potential is capable of
nicely reproducing all the DFT results, capturing both the values and the signs of the binding energies accurately.
This is a remarkable outcome,
since machine-learning energies for defects
are much more challenging than for bulk systems~\cite{PhysRevX.8.041048,PhysRevMaterials.2.013808,PhysRevB.100.144105}.
Considering the limited descriptive capacity of the EAM model, the performance of the EAM potential is acceptable and somehow unexpected.
The noticeable disparities between EAM and the DFT results are evident in the description of the binding energies of vacancy clusters and formation energy of antisite defect Ni$_{\rm Al}$. The EAM potential inaccurately predicts a negative formation energy of Ni$_{\rm Al}$.
The MTP-calculated formation energy of a vacancy in bulk Ni
is 1.425 eV, in reasonable agreement with the scattered experimental data (1.4$\sim$1.74 eV)~\cite{glazkovFormationPointDefects1987,nanaoStudiesDefectsThermal1977,wolffVacancyFormationNickel1997}.
The monovacancy in bulk Ni$_3$Al demonstrates a slightly higher formation energy. The MTP-predicted formation energies for a Ni vacancy and an Al vacancy are 1.588 eV and 1.710 eV, respectively, suggesting the relatively easier formation of a Ni vacancy in bulk Ni$_3$Al. Experimental differentiation between Ni and Al vacancies is challenging, leading to an averaged formation energy of 1.6$\pm$0.2 eV \cite{wangStudyVacanciesIntermetallic1984a}.
It is important to note that the MTP model inherits the limitations of the underlying DFT, which demonstrates inaccuracies in predicting surface energy when using the PBE functional~\cite{mattssonElectronicSurfaceError2008,nandiEfficacySurfaceError2010,nazarovVacancyFormationEnergies2012,michaelidesIntroductionTheoryCrystalline2013,medasaniVacancyFormationEnergies2015}.
This could be linked to the precision of the predicted formation and binding energies of vacancies.

Regarding the binding energies of vacancy clusters,
it is interestingly found that the divacnacies in the first-nearest neighbor ($1NN$) in the bulk Ni exhibit a weak attractive interaction. However, as the distance between the two Ni vacancies increases, the interaction becomes repulsive.
In the bulk Ni$_3$Al system,
an opposite trend is observed for the two Ni vacancies.
The two Al vacancies in the $1NN$ demonstrate a weak attractive interaction, whereas an Al vacancy and a Ni vacancy exhibit repulsive interactions up to the second-nearest neighbor.

\begin{figure}[H]
	\begin{center}
 \includegraphics[width=0.95\textwidth]{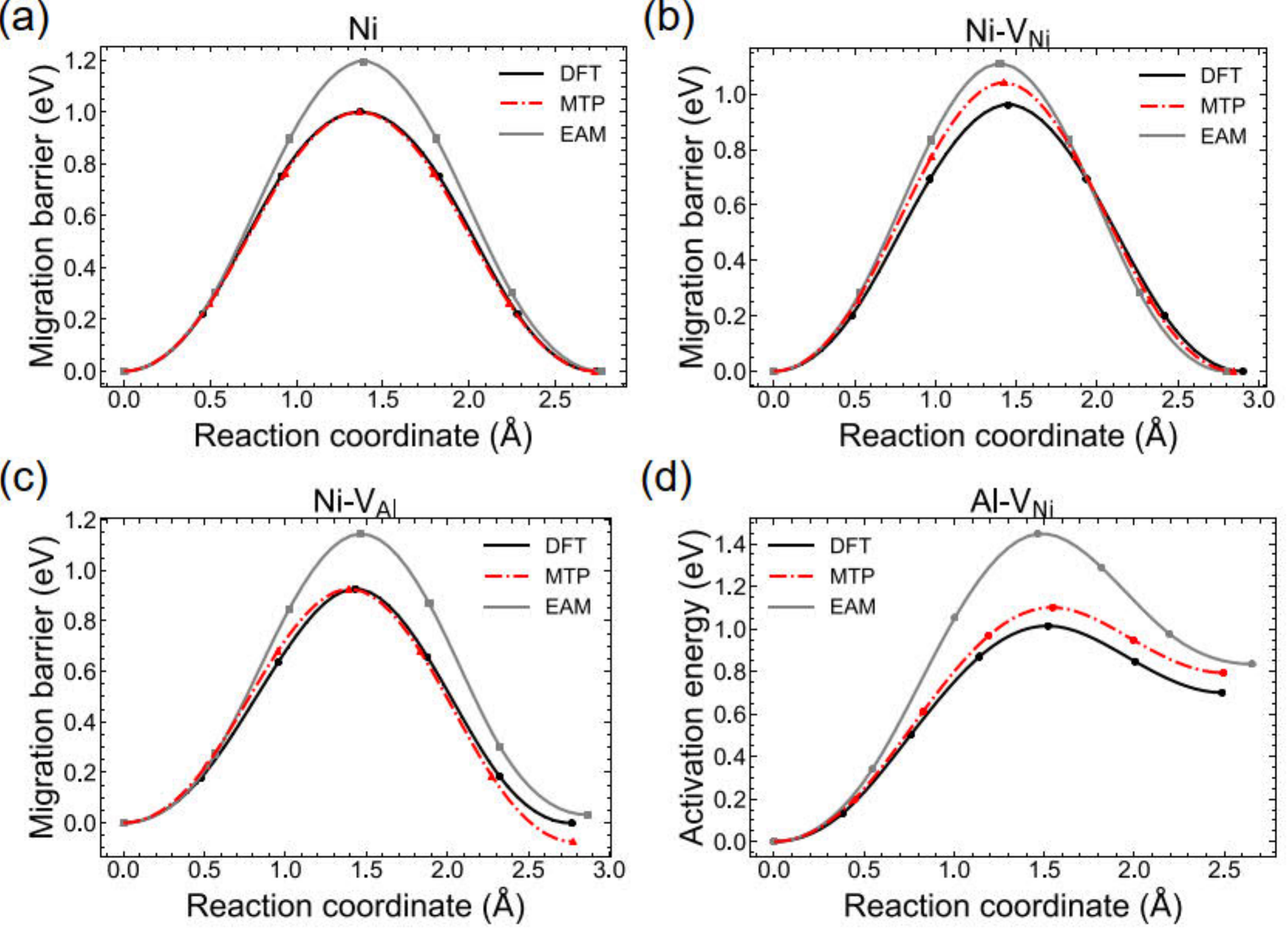}
	\end{center}
	\caption{(a) The MEP of a Ni vacancy diffusion in bulk Ni. (b) The MEP of a Ni atom diffusing to the nearest Ni vacancy in bulk Ni$_3$Al. (c) The MEP of a Ni atom diffusing to the nearest Al vacancy in bulk Ni$_3$Al. (d) The MEP of a Al atom diffusing to the nearest Ni vacancy in bulk Ni$_3$Al. The DFT, MTP, and EAM predicted results are shown in gray-solid lines, red-dashed lines, and black-solid lines, respectively.
	}
	\label{fig:meps}
\end{figure}

Figure~\ref{fig:meps} illustrates the predicted MEPs for a vacancy migration to its $1NN$ site in both bulk Ni and bulk Ni$_3$Al systems using DFT, MTP, and EAM methods.
It is evident that the MTP potential aligns well with the DFT predictions, whereas the EAM potential overestimates all the migration energy barriers as compared to DFT.
However, the EAM-predicted migration energy barrier of a Ni vacancy is 1.195 eV, which is closer to the experimental values (see Table~\ref{tab:pointdefects}). This is because the EAM potential developed by Mishin~\cite{purjapunDevelopmentInteratomicPotential2009} was obtained by deliberately fitting the migration energy barrier. Overall, the DFT-predicted migration energy barrier of a vacancy in Ni is greater than that in Ni$_3$Al, indicating a faster vacancy diffusion in Ni$_3$Al.

\begin{figure}[H]
	\begin{center}
		\includegraphics[width=0.95\textwidth]{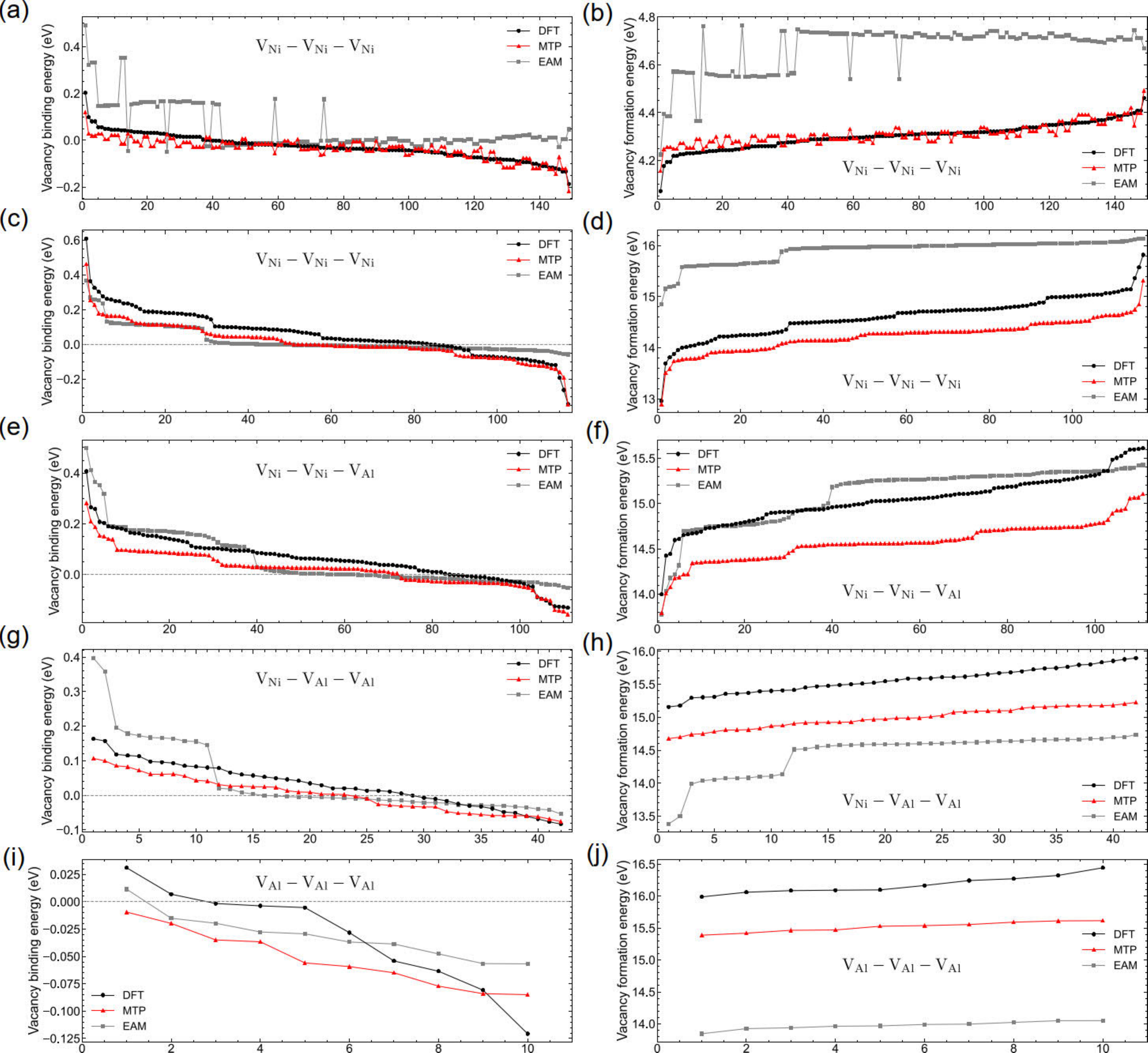}
	\end{center}
	\caption{(a)(b) Binding and formation energies of trivacancies in Ni. (c)-(j). Binding and formation energies of trivacancies in Ni$_3$Al. The horizontal axis indicates the ID of the trivacancies configurations, which are ordered by DFT-calculated formation energies.
	}
	\label{fig:trivac}
\end{figure}

We also computed the binding and formation energies of trivacancies in both Ni and Ni$_3$Al systems using DFT, MTP, and EAM methods.
In the case of trivacancies in Ni, we meticulously analyzed 144 potential symmetrically nonequivalent configurations within a $3\times3\times3$ supercell. For the trivacancies in Ni$_3$Al, we extensively explored a total of 280 energetically nondegenerate configurations within a $3\times3\times3$ supercell.
The results are compiled in Fig.~\ref{fig:trivac}.
One can see that
the MTP potential is able to capture the overall trend observed in the DFT results.
In the case of trivacancies in Ni, the MTP potential correctly predicted
the energetically most favorable trivacancies,
which form a triangle in the \{111\} plane.
The second energetically most stable trivacancies
forming a line along the $\langle111\rangle$ direction
were also accurately reproduced by MTP.
The least favorable trivacancies were found to
form a line the $\langle001\rangle$ direction.
In the case of trivacancies in Ni$_3$Al, the developed MTP potential was also capable of accurately predicting the most and least favorable configurations.
Nevertheless, given the accuracy of the MTP model, distinguishing the relative stability of the two trivacancies becomes highly challenging when their formation energy difference is less than 0.03 eV. This discrepancy contributes to variations in the relative stability of different trivacancies between the MTP and DFT predictions.
It is also notable that the current MTP potential exhibits  a larger error for trivacancies that involve Al vacancies, which is likely due to insufficient sampling of phase space encompassing the Al vacancies. This underscores the need for improving the training set in future work.
In contrast to the MTP potential, the EAM potential can only correctly predict the energetically most favorable configuration of trivacancies and  typically shows a larger deviation with the DFT results.

\begin{figure}[H]
	\begin{center}
		\includegraphics[width=0.7\textwidth]{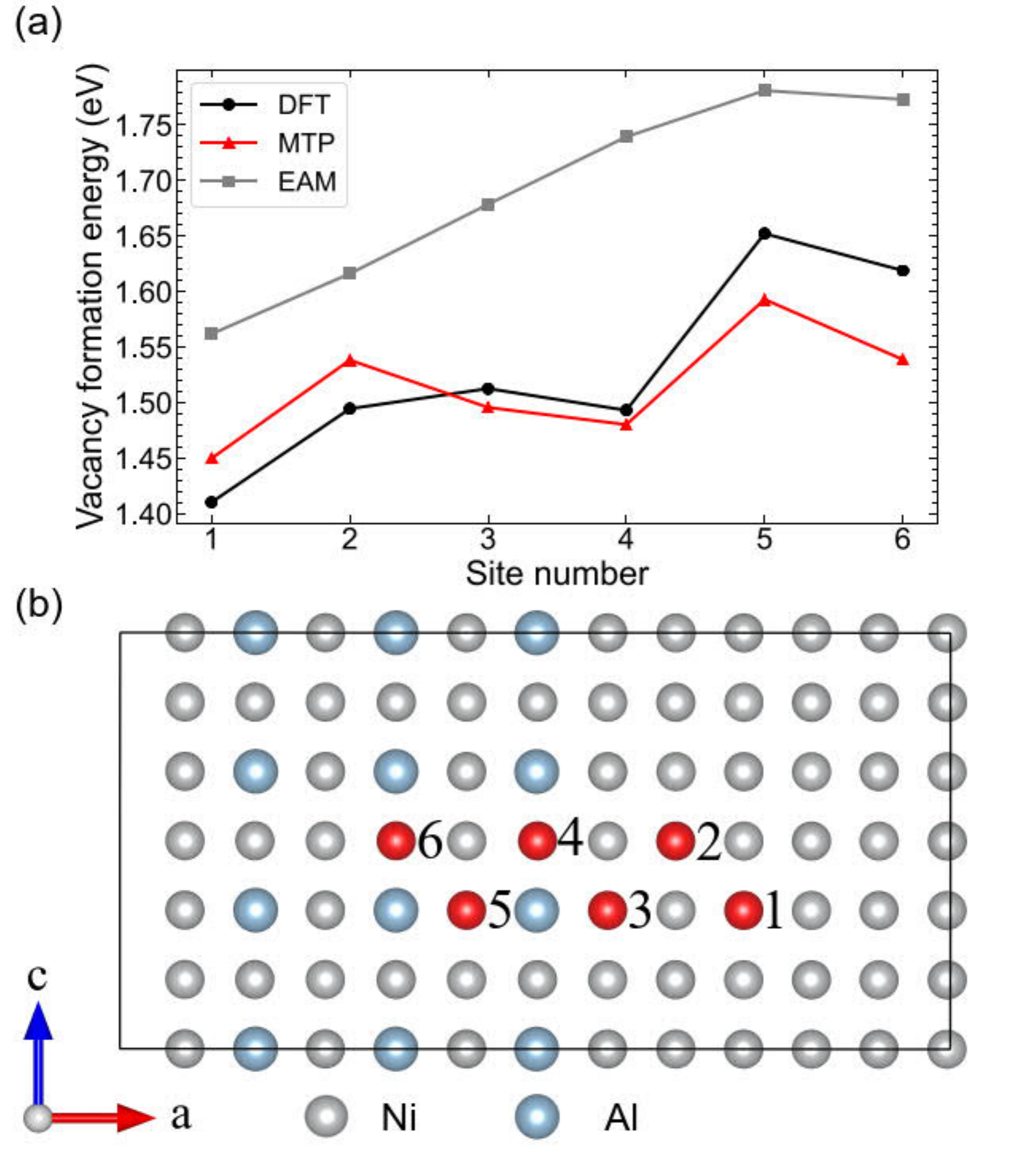}
	\end{center}
	\caption{(a) Formation energies of vacancies at the Ni-Ni$_3$Al interface boundary predicted by DFT, MTP, and EAM methods. (b) The positions of the vacancies (red balls). The site number is indicated. The blue and gray balls represent Al and Ni atoms, respectively.
	}
	\label{fig:infvac}
\end{figure}

Furthermore, we evaluated formation energies of
vacancies at the Ni-Ni$_3$Al interface boundary.
As depicted in Fig.~\ref{fig:infvac},
all the three (DFT, MTP, and EAM) methods predict that the Ni vacancy at the site 1 is the most favorable.
The MTP potential can effectively reproduce the DFT-derived relative stability of various vacancies, while the EAM potential significantly overestimates the formation energies. Again, within the accuracy, the current MTP struggles to correctly determine the order of stabilities for the vacancies at sites 2 and 3.

\subsection{Ground-state configuration of vacancy clusters in Ni}

Vacancy clusters in metals can be formed under various working conditions, including tensile or shear loads, irradiation damage, and so on~\cite{sunPrecipitationStrengtheningAluminum2019,kraftmakherEquilibriumVacanciesThermophysical1998,kiritaniSimilarityDifferenceFcc2000,nagumoPredominantRoleStraininduced2019,parkQuantitativeStudyVacancy1983}, which therefore necessitates special investigation.
It is widely accepted that in FCC metals, vacancy clusters prefer stacking fault tetragonal configurations and void configurations.
While numerous studies explored these configurations using DFT~\cite{nivaccluster1,megchicheFirstPrinciplesCalculations2010a} or semiempirical potentials~\cite{huangEffectsAppliedMechanical2021,lounisStabilityVacancyClusters2016,vasilyevSmallVacancyClusters1985,huangEffectsAppliedMechanical2021,chakrabortyCrinducedFastVacancy2017,aidhyPointDefectEvolution2015}, the stable configurations of vacancy clusters remain elusive owing to the high computational cost of DFT and the limited accuracy of semiempirical potentials.
To address this issue, here we attempted to examine the stability and ground-state configurations of vacancy clusters in nickel using our developed MTP potential.

\begin{figure}[H]
    \begin{center}
        \includegraphics[width=0.95\textwidth]{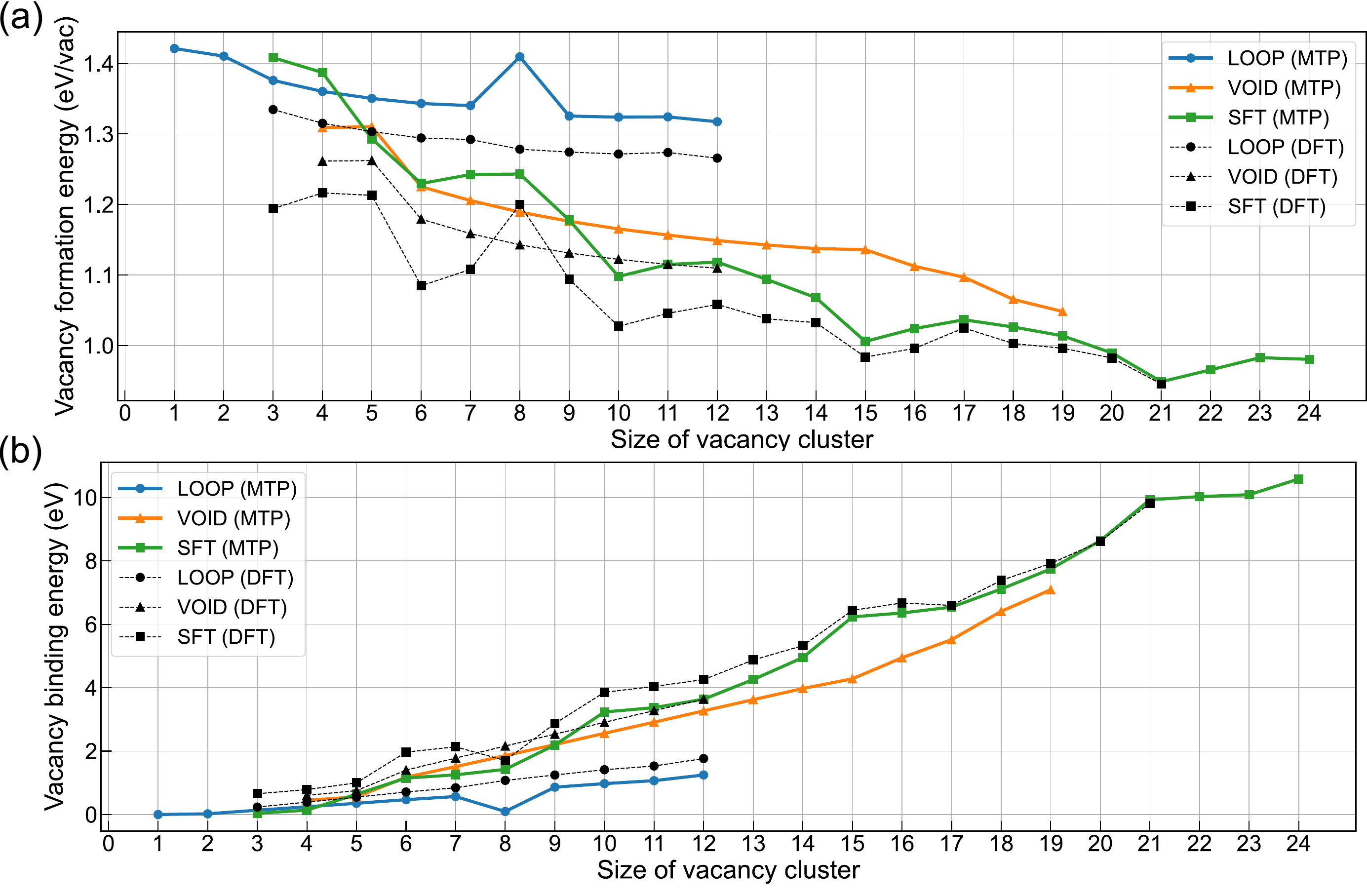}
    \end{center}
    \caption{(a) Formation energies and (b) binding energies of the vacancy clusters of LOOP, SFT, and VOID types predicted by DFT and MTP methods.}
    \label{fig:vac555}
\end{figure}

We first assessed the accuracy of the MTP potential using  $5\times5\times5$ supercell, a size affordable for DFT calculations.
Figure~\ref{fig:vac555} presents the formation energies and
binding energies of the vacancy clusters of loop, SFT, and void types predicted by DFT and MTP methods.
One can see that as compared to the DFT results,
the MTP predictions generally overestimate the formation energy per vacancy, but underestimate the binding energy of the vacancy clusters.
For vacancy clusters containing fewer than three vacancies, the current MTP potential fails to predict the correct relative order of binding and formation energies across the three types of vacancy clusters.
We would like to note that reproducing the formation and binding energy of large defect clusters using machine-learned interatomic potentials is highly challenging owing to the potential accumulation of prediction errors with increasing system sizes. Nevertheless, the overall trend of binding energy remains consistent with the anticipated behavior.

Then, we turned to a larger $8\times8\times8$ supercell consisting of 2048 atoms to determine the ground-state configuration of the vacancy clusters at their equilibrium volume and at 0 K.
The vacancy loop with a triangular shape, void, and SFT configurations were considered. For the vacancies loop and void configurations, we conducted an exhaustive search considering all possible configurations.
To identify the most stable configuration for the SFT and other possible configurations, we conducted multiple molecular dynamics annealing simulations.

\begin{figure}[H]
	\begin{center}
		\includegraphics[width=0.95\textwidth]{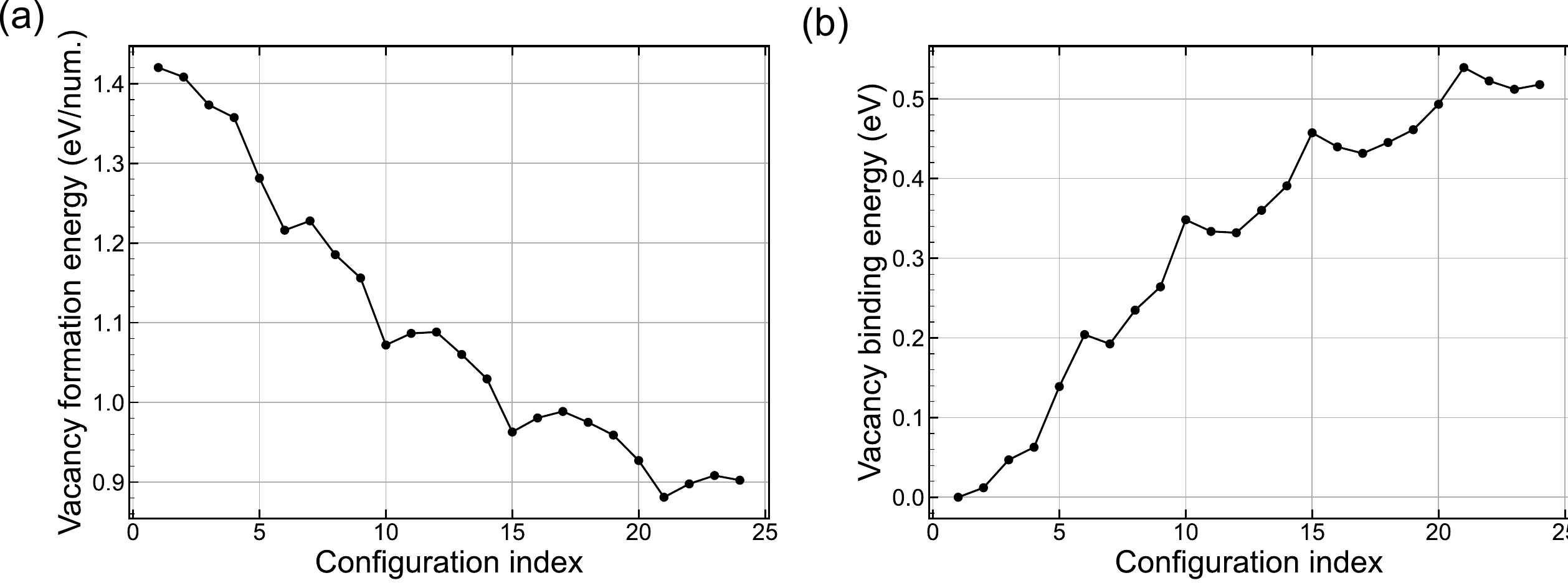}
	\end{center}
	\caption{(a) Formation energies and (b) binding energies of vacancy clusters of different sizes in an $8\times8\times8$ supercell of Ni predicted by MTP.
	}
	\label{fig:vaccls}
\end{figure}

\begin{figure}[H]
	\begin{center}
		\includegraphics[width=0.7\textwidth]{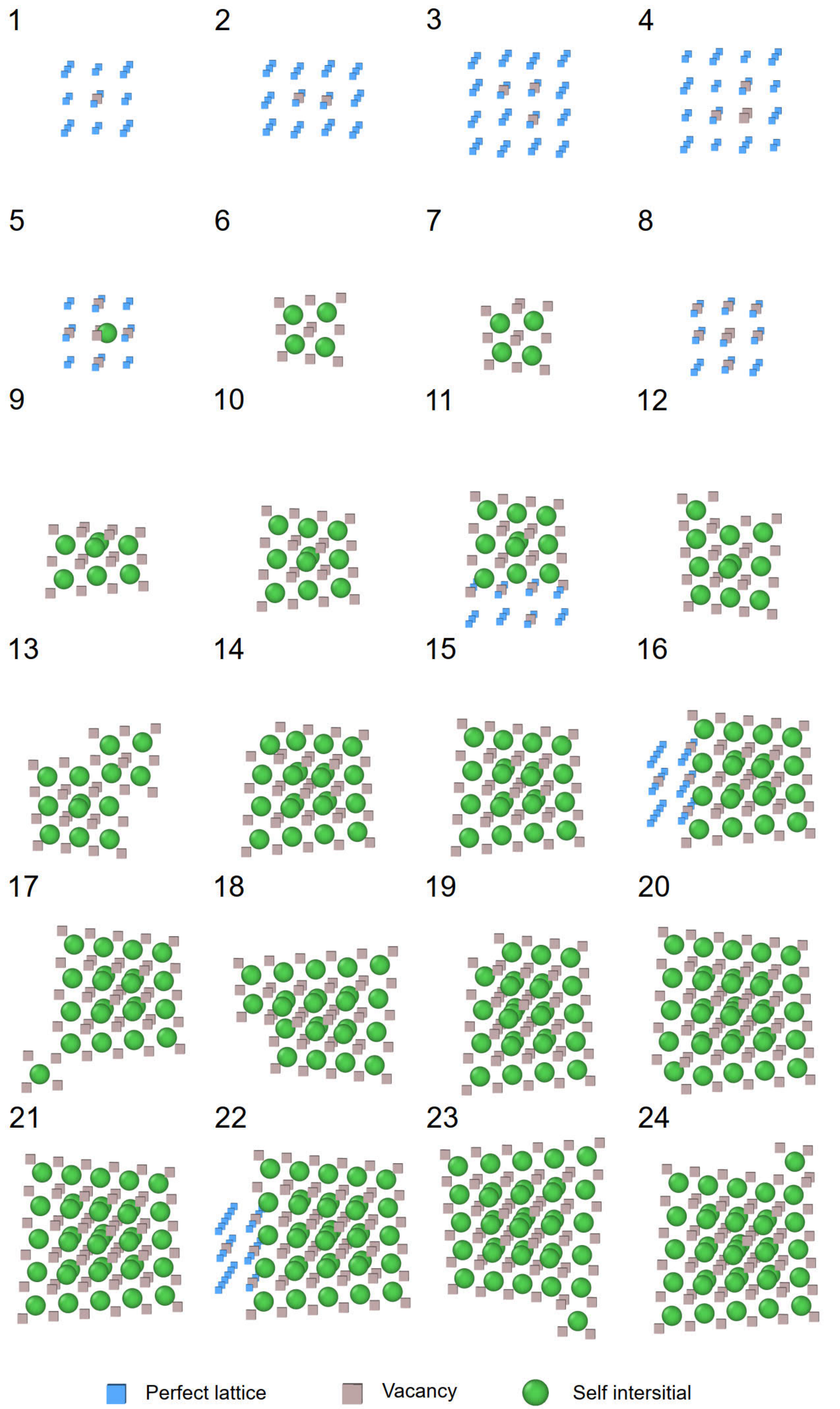}
	\end{center}
	\caption{Configurations of vacancies from monovacancy to vacancy cluster consisting of up to 24 vacancies in Ni. The green balls represent self interstitial atoms within the perfect lattice, the gray cubes denote the vacancy sites, and blue cubes represent the positions of atoms in the perfect lattice. Note that only the environment containing the defects is displayed.
	}
	\label{fig:vacimg}
\end{figure}

Figure~\ref{fig:vaccls} shows the MTP-predicted formation and binding energies as a function of the size of vacancy clusters in
an $8\times8\times8$ supercell of Ni.
The most stable configurations of vacancy clusters for each size are depicted in Fig.~\ref{fig:vacimg}.
For simplicity, in the following, we denote the cluster of $n$ vacancies as $n$v.
It can be seen that
the 3v cluster stabilizes the loop configuration,
while 4v and 8v clusters favor the void configuration.
The 5v cluster forms a 6v void configuration with an self interstitial atom (SIA) near the center.
For the remaining vacancy clusters,
the SFT configuration is found to be the most stable.
Specifically, the ideal SFT configuration is attained for the 6v, 10v, 15v, and 21v clusters, known as magic numbers.
For other vacancy clusters whose size does not align with these magic numbers, the SFT configuration is imperfect, with a neighboring vacancy.

\subsection{Diffusion behavior of vacancy clusters in Ni}

\begin{figure}[H]
	\begin{center}
		\includegraphics[width=0.8\textwidth]{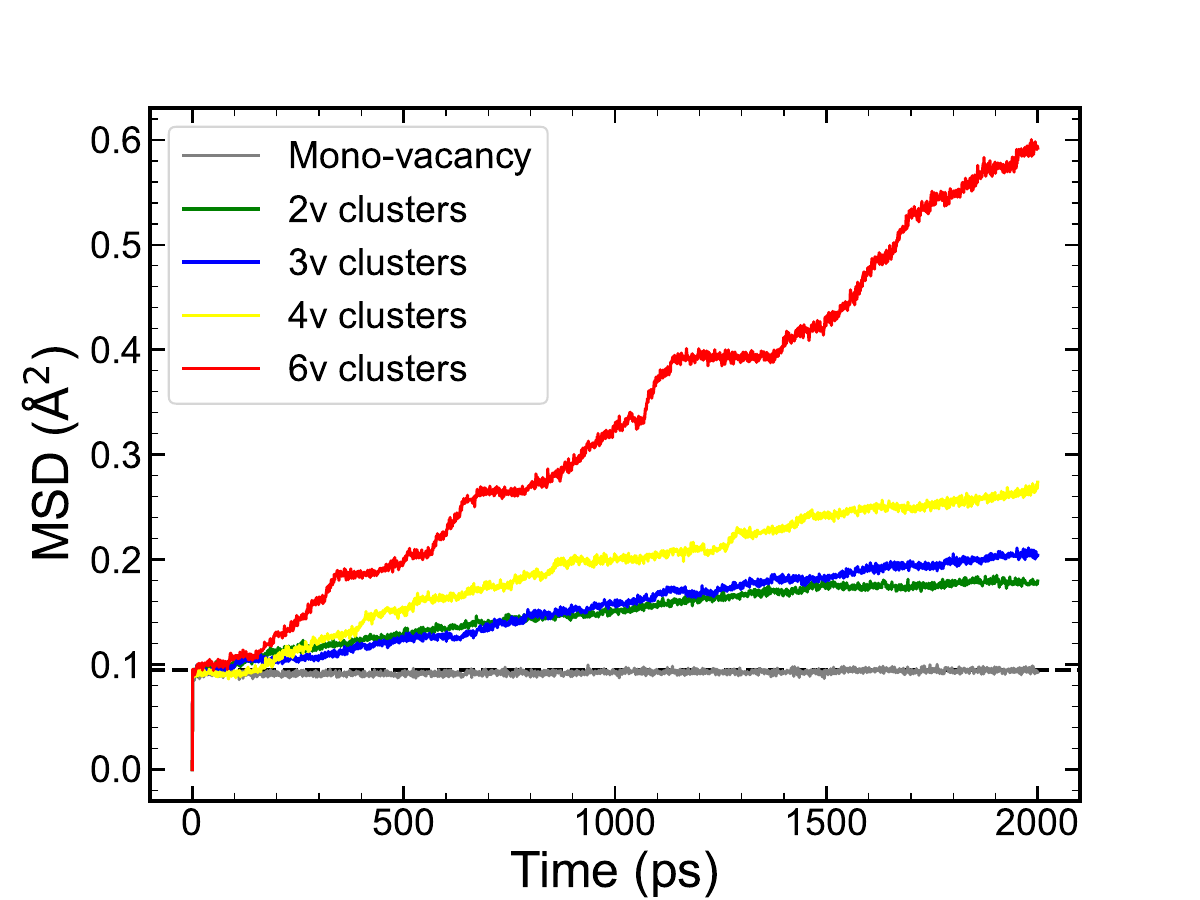}
	\end{center}
	\caption{MTP predicted mean squared displacements in the defective Ni system containing various type of vacancy clusters at a temperature of 800 K.
	}
	\label{fig:msd}
\end{figure}

In addition to investigating the static energetics of vacancy clusters, we delved into the dynamic diffusion characteristics.
In order to ensure the same vacancy concentration, we constructed an $12\times12\times12$ supercell containing twelve vacancies for each type of vacancy clusters.
By doing so, we restricted ourselves to the clusters of monovacancy, divacnacies, trivacancies, quadvacancies, and hexavacancies.
In other words, within each system, there are six clusters of divacnacies, four clusters of trivacancies, three clusters of quadvacancies, and two clusters of hexavacancies.
Using our developed MTP potential we performed molecular dynamics simulations at 800 K over a duration of 2 ns and then computed the mean squared displacement (MSD) of each system. The results are shown in Fig.~\ref{fig:msd}.
We note that in the absence of vacancy diffusion, the atoms also underwent vibrational motion around their equilibrium positions, leading to a MSD of 0.1 \AA$^2$, represented by the dashed horizontal line in Fig.~\ref{fig:msd}.

We observed during the MD run that the divacnacy and hexavacancy clusters are stable, keeping the cluster size from the beginning to the end.
However, the trivacancy cluster is unstable, decomposing into a divacnacy cluster and a nearby monovacancy after several migration steps.
After about 1 ns,
all four clusters of trivacancies underwent decomposition. During this process, the decomposed monovacancy remains unchanged, whereas the decomposed divacnacy clusters exhibit behavior akin to regular divacnacy clusters.
The quadvacancy clusters are also unstable. They initially broke down into a trivacancy cluster and a monovacancy after several migration steps. Finally, these trivacancy clusters further decomposed into a divacnacy cluster and a monovacancy.
Because of the decomposition process, the ultimate slopes of the MSD curves for divacnacy, trivacancy, and quadvacancy clusters are hence nearly indistinguishable.
Under 800 K, the monovacancy is almost immobile,
with only one jump observed throughout the entire trajectory.
Interestingly, the hexavacancy clusters with the SFT structure (6v-SFT) exhibit the fastest diffusion.

We conducted an in-depth analysis to understand why the 6v-SFT cluster diffuses so rapidly compared to the other five vacancy clusters under consideration. Within the 6v-SFT clusters, there are two configurations that are mutually centrosymmetric with each other. We ran another MD simulation with a shorter dump interval, containing only one 6v-SFT cluster in the system. During this simulation, we observed frequent transitions between the two configurations of the 6v-SFT cluster. The calculated transition barrier at 0 K using the CI-NEB method is approximately 0.4 eV, surprisingly low for such transitions to occur.

% pick: 200 - 341 - 381 for the first jump    jump.ovito
% pick: 1130 - 1140 - 1150 for the second one jump2.ovito

\begin{figure}[H]
	\begin{center}
		\includegraphics[width=0.9\textwidth]{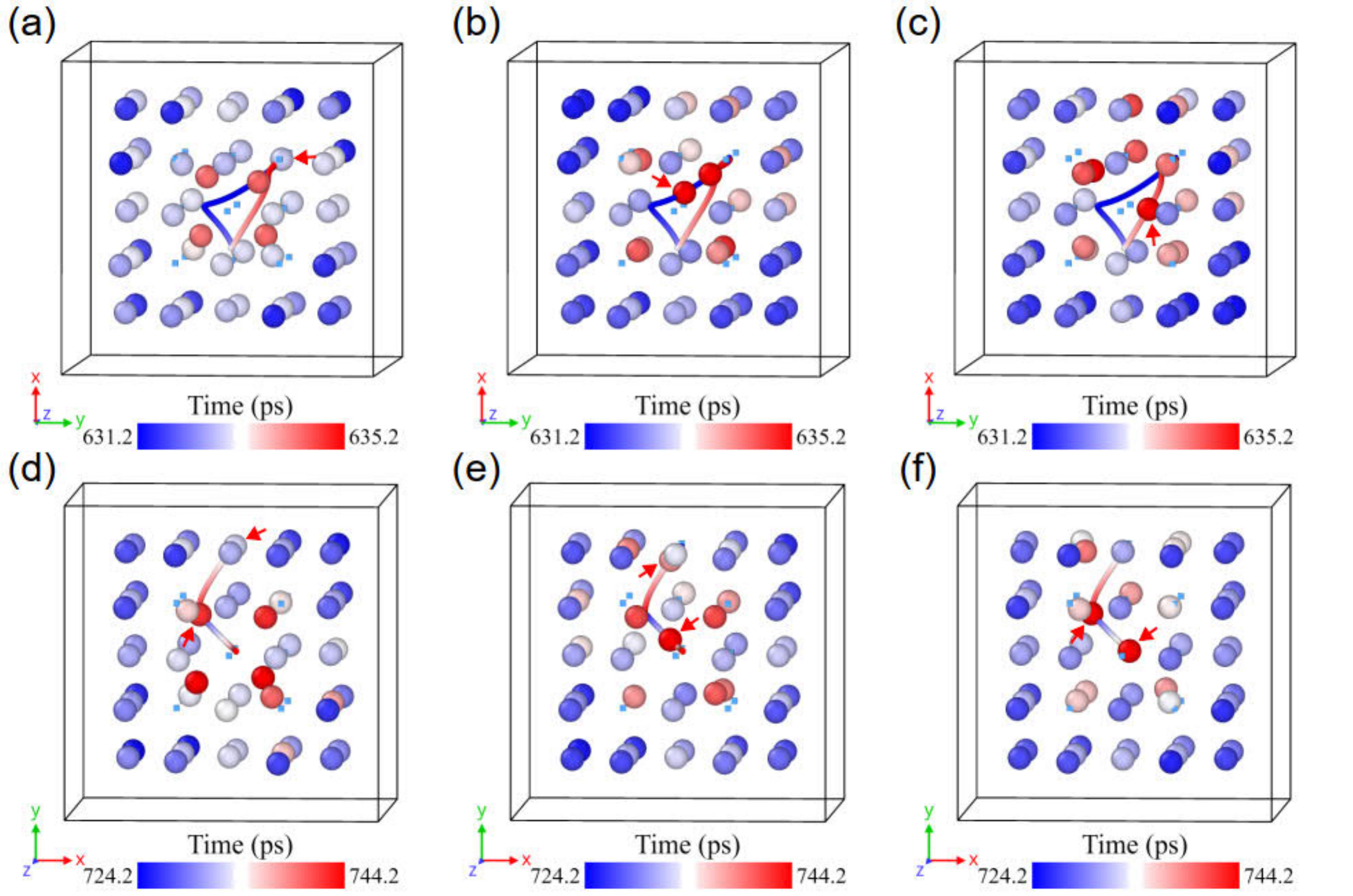}
	\end{center}
	\caption{Snapshots from a MD trajectory. Note that only atoms near the SFT cluster are shown.
 The atoms are color-coded based on their site energies predicted by our MTP potential.
 The blue cubes represent the positions of atoms in the perfect lattice.
 The trajectory of the migrating atoms (marked by red arrows) is colorized according to simulation time. (a)-(c) Snapshots of the first migration event occurring between 631.2 ps and 635.2 ps. (d)-(f) Snapshots of second migration event occurring between 724.2 ps and 744.2 ps.
	}
	\label{fig:mdsnapshot}
\end{figure}

To uncover the migration path of the 6v-SFT cluster, we applied a low pass filter with a super-Gaussian window on the trajectory to effectively filter out thermal vibrations above 10 THz. At 631.2 ps, during the transition between the two configurations of the SFT cluster, an atom underwent a rapid and long-distance migration across the core region. It moved swiftly, covering a significant distance, and then quickly returned to its original position within a few femtoseconds. The trajectory of the migrating atom is shown in Figs.~\ref{fig:mdsnapshot}(a)-~\ref{fig:mdsnapshot}(c). Notably, a hole appeared at the center of the cluster during the transition between the two symmetrically opposite configurations of the 6v-SFT, potentially aiding in the occurrence of the observed long-distance transfers.

Similarly, at 724.2 ps, a migration event similar to the one observed at 631.2 ps reoccurred. However, this time another atom took the place of the previously migrated atom, as shown in Fig.~\ref{fig:mdsnapshot}(d)-\ref{fig:mdsnapshot}(f). This caused the cluster to enter an unstable state characterized by rapid transfers of core atoms. This unstable state persisted for over 500 ps in the trajectory, corresponding to the rising section of the MSD curve in Fig.~\ref{fig:msd}. Once the cluster returned to a stable 6v-SFT configuration, the rapid migrations ceased, rendering the cluster immobile once more. Following this process, the 117 atoms of the 6v-SFT cluster including the initial SFT core atoms, migrated approximately 17 \AA.
For enhanced visualization of the migration process of the 6v-SFT cluster, we refer to the Supplemental Material~\cite{SM}.

\subsection{Description of the energetics of planar defects by MLIP}

Having assessed the accuracy of our developed MLIP in describing point defects, we now turn to evaluating its performance in the description of planar defects such as generalized stacking faults (GSF). The GSF was introduced by Vitek~\cite{vitekIntrinsicStackingFaults1968} to describe the energy variations that occur when half of a crystal undergoes shear displacement across a glide plane in specific directions. It provides valuable insights into the behavior of intrinsic stacking faults and the deformation mechanisms of materials~\cite{benyoucefStackingFaultEnergy1995,hullDislocationsFacecenteredCubic2011}.
Accurate modeling of stacking faults is crucial for understanding the strength and plasticity of materials,
since the intrinsic stacking fault energies and distances of these faults are critical parameters for modeling dislocations~\cite{hirthTheoryDislocations2nd1983}.

Employing our developed MTP potential, we computed the generalized stacking fault energy (GSFE) for the densely packed (111) plane in both Ni and Ni$_3$Al.
To model the stacking faults,
the shear alias method~\cite{ogataIdealPureShear2002} was employed using a six-layer supercell.
The GSFE was calculated as
\begin{equation}
    \gamma_{GSF} = (E_{GSF} - E_0) / A,
\end{equation}
where $E_{GSF}$ and $E_0$ are total energies of the supercell  with or without the stacking fault, respectively, and $A$ is the area of the slip plane within the supercell.
For determining the GSFE on the (111) plane, a $16 \times 26$ grid was utilized for DFT calculations, while a dense $120 \times 120$ grid was applied for MTP and EAM calculations.

\begin{figure}[H]
	\begin{center}
		\includegraphics[width=0.9\textwidth]{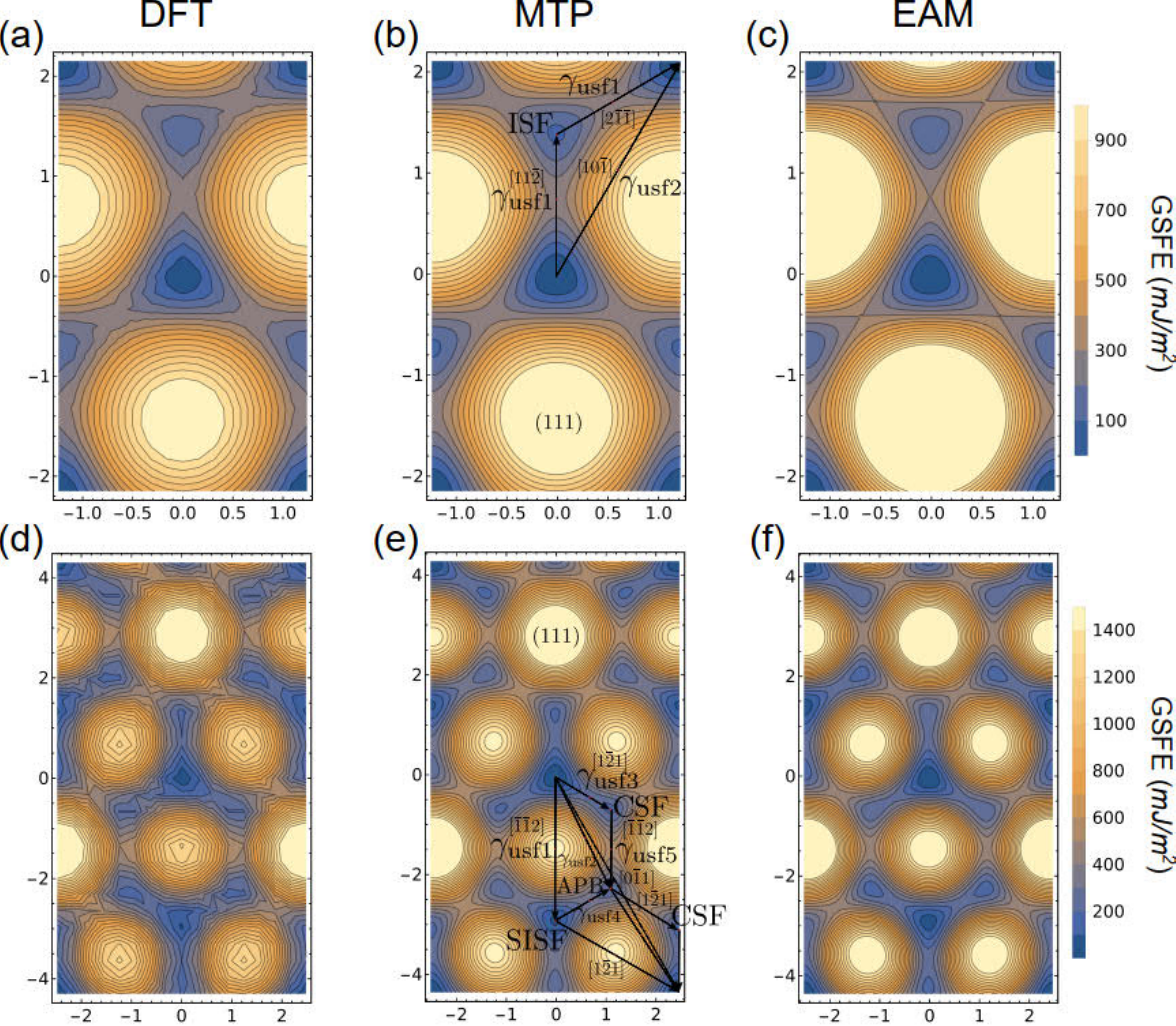}
	\end{center}
	\caption{(a)-(c) Contour plot of generalized stacking fault energy of the Ni (111) plane predicted by DFT, MTP, and EAM, respectively. (d)-(f) Contour plot of generalized stacking fault energy of the Ni$_3$Al (111) plane predicted by DFT, MTP, and EAM, respectively. The $x$ and $y$ axis denote the displacements from initial perfect lattice along the $[1\bar{1}0]$ and $[11\bar{2}]$ directions, respectively.
 The color coding indicates the generalized stacking fault energy in unit of mJ/m$^2$.
	}
	\label{fig:gsfmesh}
\end{figure}

Figure~\ref{fig:gsfmesh} presents the GSFEs of the (111) plane of Ni and Ni$_3$Al predicted by DFT, MTP, and EAM methods.
The values of various stacking fault energies are summarized in Table~\ref{tab:gsfvalues}. Overall, the MTP predictions demonstrate better agreement with the DFT results than the EAM predictions. Given the scattered nature of both the calculated and experimental literature data, our predictions are in line with them. The MTP prediction errors are negligible in the low-energy region, but tend to increase in the high-energy region.
This is caused by the inefficient sampling of these energetically unfavorable configurations from unbiased molecular dynamics simulations.

\begin{table}[H]
	\begin{center}
		\caption{Stacking fault energies of the (111) plane of Ni and Ni$_3$Al predicted by DFT, MTP, and EAM methods.
  The specific definition of each stacking fault energy has been provided in Fig.~\ref{fig:gsfmesh}.
  The results are compared to other calculations as well as experiments. All energies are in unit of mJ/m$^2$.}
		\label{tab:gsfvalues}
		\begin{tabular}{c c | c c c c c }
  \hline
			{} & {} & DFT & MTP & EAM & Other calculations & Experiments \\
			\hline
			\multirow{3}{2.5em}{\rm Ni}
			   & $\gamma_\text{ISF}$   & 133 & 115 & 139 & 117.70 & 120$\sim$130~\cite{carterStackingfaultEnergyNickel1977}  \\
			{} & $\gamma_\text{usf1}$ & 263 & 276 & 297 & 283~\cite{huReactiveElementsDependence2020},273~\cite{XIA2022104183},278~\cite{YANG2020109682} & {}  \\
			{} & $\gamma_\text{usf2}$ & 713 & 688 & 974 & 783~\cite{huReactiveElementsDependence2020}  & {} \\
			\hline
			\multirow{7}{2.5em}{Ni$_3$Al}
			   & $\gamma_\text{CSF}$ & 210 & 196 & 238 & 205~\cite{YU201238},208~\cite{XIA2022104183},225~\cite{mryasovSuperdislocationCoreStructure2002},249~\cite{HU2020155799} & 235$\pm$45~\cite{karnthalerInfluenceFaultEnergies1996}  \\
			{} & $\gamma_\text{APB}$ & 163 & 235 & 206 & 180~\cite{YU201238},198~\cite{XIA2022104183},259~\cite{huReactiveElementsDependence2020} & 175$\pm$15~\cite{karnthalerInfluenceFaultEnergies1996} \\
			{} & $\gamma_\text{SISF}$ & 55 & 73 & 41 & 21~\cite{XIA2022104183},47~\cite{huReactiveElementsDependence2020},75~\cite{YU201238},80~\cite{mryasovSuperdislocationCoreStructure2002} & 6$\pm$5~\cite{karnthalerInfluenceFaultEnergies1996},35~\cite{knowlesSuperlatticeStackingFault2003} \\
			{} & $\gamma_\text{usf1}$ & 1324 & 1291 & 1618 & 1332~\cite{XIA2022104183},1368~\cite{YU201238},1421~\cite{huReactiveElementsDependence2020} & {}  \\
			{} & $\gamma_\text{usf2}$ & 808 & 788 & 948 & 778~\cite{YU201238},791~\cite{XIA2022104183},830~\cite{huReactiveElementsDependence2020} & {}  \\
			{} & $\gamma_\text{usf3}$ & 232 & 248 & 252 & 227~\cite{XIA2022104183},254~\cite{YU201238} & {} \\
			{} & $\gamma_\text{usf4}$ & 214 & 299 & 257 & {} & {} \\
			{} & $\gamma_\text{usf5}$ & 522 & 524 & 610 & {} & {} \\
			\hline
		\end{tabular}
	\end{center}
\end{table}

Following the minimal slide path as depicted in Fig.~\ref{fig:gsfmesh}, our results clearly demonstrated that in bulk Ni, a $\frac{1}{2}[1\bar{1}0]$ dislocation decomposes into two Schockley partial dislocations $\frac{1}{6}[2\bar{1}\bar{1}]$ and $\frac{1}{6}[1\bar{2}1]$ plus an intrinsic stacking fault (ISF).
In the case of Ni$_3$Al, a $[1\bar{1}0]$ super dislocation first decomposes into two $\frac{1}{2}[1\bar{1}0]$ dislocations plus an antiphase boundary (APB). Then, each $\frac{1}{2}[1\bar{1}0]$ decomposes into two Schockley partial dislocations $\frac{1}{6}[2\bar{1}\bar{1}]$ and $\frac{1}{6}[1\bar{2}1]$ plus a complex stacking fault.
Our results are consistent with the discussions presented in the literature~\cite{raePrimaryCreepSingle2007,leverantCreepPrecipitationhardenedNickelbase1973,matanCreepCMSX4Superalloy1999,knowlesSuperlatticeStackingFault2003,breidiFirstprinciplesModelingSuperlattice2018}.

It is worth mentioning that the antiphase boundary (APB) in Ni$_3$Al, as shown in Fig.~\ref{fig:gsfmesh}, exhibits a slight deviation from its geometrically ideal position after structural relaxation.
For this reason, we employed the minimal energy point of the APB instead of the geometrically ideal one to determine  the stacking fault energies. As a result, our computed value of $\gamma_\text{APB}$ is lower than the values reported in the literature.

\subsection{Melting point prediction by MLIP}

As a final inference test for our developed MLIP,
we computed the melting points of Ni and Ni$_3$Al
using the solid-liquid coexistence method~\cite{zhuFullyAutomatedApproach2021}.
A $10 \times 10 \times 20$ supercell consisting of 8000 atoms was employed.
The solid and liquid phases were separated by the (001) interface.
The calculated melting point of Ni is around 1597 K, which underestimates the experimental value of 1728 K~\cite{cahnBinaryAlloyPhase1991,huangThermodynamicAnalysisNi1998}.
For Ni$_3$Al, the MTP-predicted melting point is 1630 K,
in good agreement with the experimental value of 1645 K~\cite{cahnBinaryAlloyPhase1991,huangThermodynamicAnalysisNi1998}.
We note that our training set did not deliberately incorporate melting structures. It is therefore remarkable that the MTP potential we developed accurately predicts the melting points of both Ni and Ni$_3$Al, showcasing a notable extrapolation capacity.

\section{Conclusions}\label{sec:conlcusions}

In conclusion, we have proposed an effective genetic algorithm based optimization scheme
for moment tensors contractions,
which significantly reduces the number of independent moment tensor components and intermediate tensor components.
This results in almost a tenfold acceleration in speed and improved accuracy as compared to the traditional MTP model of Shapeev~\cite{mtp1}
for the basis sets with a high level of complexity.
The performance of our improved MTP model has been thoroughly assessed
by predicting the energetic and dynamical properties of various point and planar defects in Ni-Al alloys.
We found that our developed MTP model not only is capable of reproducing the fundamental physical properties of bulk Ni and Ni$_3$Al
such as the lattice constants, elastic properties, energy-volume curves, phonon dispersions, and the melting points,
but also can accurately predict the formation and binding energies of vacancy clusters, antisite defects, and stacking faults
as well as the diffusion behavior of vacancy clusters, in general surpassing the widely used semiempirical EAM potentials.
The latter perform well for the target properties used in fitting,
but tend to exhibit large errors for complex defects such as vacancy clusters and stacking faults.
Our developed MTP model also enables the identification of the ground-state configuration of vacancy clusters of differing sizes,
accurately predicting the optimal SFT configuration for the 6v, 10v, 15v, and 21v clusters.
Interestingly, we found that the 6v cluster with the SFT configuration is very stable and diffuses fastest
among the considered vacancy clusters (1v, 2v, 3v, 4v, and 6v) with the same vacancy concentration.
Furthermore, our study underscores the general challenge for MLIP models to predict the correct relative stability order
of the defects with comparable formation energies, which probably requires an accuracy of less than a meV per atom.
While more complex basis sets can achieve this high level of accuracy, it comes at the expense of efficiency.
Finally, we would like to note that the current MTP model can be systematically improved
by broadening the phase space of the training set through the active learning approach,
thereby laying the foundation for developing a general-purpose MLIP for Ni-Al alloys.

\section*{Acknowledgements}
This work is supported by
the Strategic Priority Research Program of Chinese Academy of Sciences (Grant No. XDA041040402),
the National Natural Science Foundation of China (Grants No. 52422112, No. 52188101, and No. 52201030),
the National Key R{\&}D Program of China 2021YFB3501503,
the Liaoning Province Science and Technology Planning Project (2024JH1/11700032, 2023021207-JH26/103 and RC230958),
and
the Special Projects of the Central Government in Guidance of Local Science and Technology Development (2024010859-JH6/1006).
Part of the numerical calculations in this study were carried out on the ORISE Supercomputer (Grant No. DFZX202319).

% \section*{Author contributions}
P.L. and X.-Q.C. conceived and supervised the project.
J.W. implemented the method. J.W. and P.L. performed the calculations.
Z.H., M.L., H.M., Y.C. and Y.S. participated in discussions.
P.L. and J.W. wrote the manuscript. All authors comment on the manuscript.

\section*{Data availability}
The data that support the findings of this article are openly availiable~\cite{supp}.

\bibliography{Reference}% Produces the bibliography via BibTeX.

\end{document}